\documentclass[a4paper,11pt]{article}
\pdfoutput=1 

\usepackage{jheppub} 

\usepackage{hyperref}
\usepackage{pdfpages}
\numberwithin{figure}{section}

\usepackage{epstopdf}

\usepackage{epsfig}
\usepackage{graphicx,color}

\newcommand{\be}{\begin{equation}}
\newcommand{\ee}{\end{equation}}

\newcommand{\tr}{{\rm tr}}

\newcommand{\rmq}{{\rm \bf q}}
\newcommand{\rmg}{{\rm \bf g}}
\newcommand{\rmp}{{  \gamma}}

\newcommand{\q}{{ \underline{q}}}

\newcommand{\beq}{\begin{eqnarray}}
\newcommand{\eeq}{\end{eqnarray}}

\definecolor{red}{rgb}{1,0,0}
\definecolor{gray}{rgb}{0.5,0.5,0.5}

\usepackage{hyperref}

\def\bea{\begin{eqnarray}}
\def\eea{\end{eqnarray}}

\title{Soft photon and two hard jets forward production in proton-nucleus collisions}

\author[a]{Tolga Altinoluk,}
\author[b]{N\'{e}stor Armesto,}
\author[c,d,e]{Alex Kovner,}
\author[d]{Michael Lublinsky}
\author[f,g]{and Elena Petreska}

\affiliation[a]{National Centre for Nuclear Research, 00-681 Warsaw, Poland}
\affiliation[b]{Departamento de F\'{i}�sica de Part\'{i}culas, AEFIS and IGFAE, \\Universidade de Santiago de Compostela, 15782 Santiago de Compostela, Galicia-Spain}
\affiliation[c]{Physics Department, University of Connecticut, 2152 Hillside road, Storrs, CT 06269, USA}
\affiliation[d]{Physics Department, Ben-Gurion University of the Negev, Beer Sheva 84105, Israel}
\affiliation[e]{Theoretical Physics Department, CERN, CH-1211 Geneve 23, Switzerland}
\affiliation[f]{Department of Physics and Astronomy, VU University Amsterdam, De Boelelaan 1081,\\
NL-1081 HV Amsterdam, The Netherlands}
\affiliation[g]{Nikhef, Science Park 105, NL-1098 XG Amsterdam, The Netherlands}

\emailAdd{tolga.altinoluk@ncbj.gov.pl}
\emailAdd{nestor.armesto@usc.es}
\emailAdd{kovner@phys.uconn.edu}
\emailAdd{lublinm@bgu.ac.il}
\emailAdd{petreska@nikhef.nl}

\date{\today}

\preprint{CERN-TH-2018-021, Nikhef 2018-006}

\abstract{
We calculate the cross section for production of a soft photon and two hard jets in  the forward rapidity region in proton-nucleus collisions at high energies. The calculation is performed within the hybrid formalism.  The hardness of the final particles is defined with respect to the saturation scale of the nucleus. We consider both the correlation limit of small momentum imbalance and the dilute target limit where the momentum imbalance is of the order of the hardness of the jets. The results depend on the first two transverse-momentum-dependent (TMD) gluon distributions of the nucleus.
}

\begin{document}
\maketitle
\flushbottom
\allowdisplaybreaks


\section{Introduction}

Photon production in nuclear collisions is a very interesting process for several reasons. First, it can be used as a tool to constrain nuclear parton density functions (nPDFs) \cite{Arleo:2011gc} in the standard collinear framework. Second, it could be sensitive to deviations from collinear factorisation that are expected at high energies or small momentum fractions $x$, particularly to non-linear effects proposed long ago \cite{GLRMQ} and subsequently developed in \cite{Mueller,balitsky,Kovchegov,JIMWLK,cgc} leading to the Color Glass Condensate framework (CGC). Finally, prompt thermal photons constitute a probe of the characteristics and dynamics of the medium created in heavy-ion collisions \cite{Stankus:2005eq}. In this respect, accurate calculations of non-thermal prompt photon production are most interesting as they constitute the background for the probe.

In the collinear framework, next-to-leading order (NLO) calculations have been available for several decades \cite{collph}. Calculations within the $k_T$-factorization scheme are also available \cite{Lipatov:2005wk} although they are still restricted to a regime in which both participating hadrons can be considered as dilute partonic objects. In the dense non-linear regime, only recently an NLO calculation has appeared \cite{Benic:2016uku} that is adequate for central rapidities in proton-nucleus (p-A) collisions. On the other hand, calculations for forward rapidities (in the proton direction), a situation in which the so-called hybrid formalism \cite{Dumitru:2005gt}
is suitable, would be very useful. In this kinematic region, where the proton can be described using collinear PDFs while the nucleus has to be treated as a dense object, the partonic structure of the nucleus can be probed at small $x$ and experiments at the LHC, particularly LHCb and future upgrades of ALICE, should be able to perform the relevant measurements.

Within the hybrid formalism, NLO corrections to light \cite{Elastic_vs_Inelastic,nlohybridpart,nlohybridpart2} and heavy \cite{Altinoluk:2015vax} hadron production have been computed, but only leading-order expressions are currently available for photon production \cite{Gelis:2002ki}. In this work we  compute the NLO corrections which are leading  in a specific kinematic situation that allows for simple analytic expressions, shows sensitivity to saturation effects and may  be experimentally accessible.
In particular we calculate the cross section for the production of a soft photon and two hard jets in forward p-A collisions, in the small-$x$ limit. We concentrate on the process:
\begin{equation}
  p (p_p) + A (p_A) \to \gamma (q_1) + g (q_2)+ q (q_3) + X\,,
\end{equation}
with the transverse momentum of the photon smaller than or of order, and the transverse momenta of the jets much larger than, the saturation scale of the nucleus.\footnote{The proton is moving in the "+" direction and the nucleus in the "--" direction. The four-momentum of the incoming quark from the proton is denoted $(p^+,p^-,p)$.} A somewhat related calculation of photon plus two jet production, but tailored to the central rapidity region and thus dominated by different diagrams, can be found in \cite{Benic:2017znu}.

The plan of the paper is as follows. In the next Section we present the calculation of the cross section using  the wave function approach to the CGC. This approach was used by some of us previously to compute NLO corrections to particle production and to small-$x$ evolution \cite{Lublinsky:2016meo}. Then, in Section \ref{Section:Back-to-back} we consider  the limit of almost back-to-back jets, and in Section \ref{Section:Dilute target} the limit of a dilute target.
Finally, in Section~\ref{Section:Discussion} we discuss our results. The details of the calculation of the wave function are presented in Appendices \ref{Section:AppendixA} and \ref{Section:AppendixB}.

\section{The wave function approach}
\label{Section:WaveFunctional}
The dominant contribution to photon production in p-A scattering comes from the photon emission off the projectile quarks that propagate through the strong color field of the target. At NLO, of course, there are additional contributions due to splitting of the projectile gluon into a quark-antiquark pair. However this process gives negligible contribution in the kinematics we are considering: emission of a soft photon and two hard jets, since a  splitting of a gluon into a $q\bar q$ pair that subsequently emits a photon would mostly result in a photon radiation which is collinear to either the $q$ or the $\bar q$. 

We note that there are also non negligible contributions coming from collinear two-parton densities in the incoming proton, i.e., one quark emitting a photon and going through the target, while the other parton, quark or gluon, going through the target independently and producing the second jet. Analysis of this type of processes is a recent development within the the hybrid model approach \cite{Kovner:2017ssr}.
In the present paper we will not consider this kind of contributions, and this limitation of the calculation has to be kept in mind.

 Our primary interest in this paper will be therefore in the projectile quark state, which of course has to be dressed by both gluon and photon radiation at NLO.

\subsection{Dressed quark state}

The dressed quark state with $+$-momentum $p^+$, vanishing transverse momenta, spin $s$ and color $\alpha$  can be written, in full momentum space, in terms of the bare states  as\footnote{There is also an instantaneous quark contribution in the dressed quark state. However, in our kinematics this contribution is suppressed by the large momenta of the jets. Therefore, we neglect it throughout the paper.}
\beq
&&\left| (\rmq) [p^+,0]^\alpha_s\right\rangle_D= 
A^q\left| (\rmq) [p^+,0]_s^\alpha\right\rangle_0\nonumber \\ 
&+&
A^{q\gamma} \; g_e\sum_{s',\lambda} \int \frac{dk_1^+}{2\pi} \frac{d^2k_1}{(2\pi)^2} \; 
F_{(\rmq\rmp)}^{(1)}\left[ (\rmp)[k_1^+,k_1]^\lambda, (\rmq)[p^+-k_1^+,-k_1]_{ss'}\right]\nonumber\\
&&
\hspace{6cm}
\times
\left| (\rmq) \left[p^+-k_1^+,-k_1\right]_{s'}^\alpha ; (\rmp) [k_1^+,k_1]_\lambda \right\rangle_0 \nonumber\\
&+&A^{qg} \; g_s\sum_{s',\eta} \int \frac{dk_2^+}{2\pi}\frac{d^2k_2}{(2\pi)^2} \, t^c_{\alpha\beta} \; 
F^{(1)}_{(\rmq\rmg)}\left[ (\rmg)[k_2^+,k_2]^\eta, (\rmq)[p^+-k_2^+,-k_2]_{ss'}\right] \nonumber\\
&&
\hspace{6cm}
\times
\left| (\rmq)\left[p^+-k_2^+,-k_2 \right]_{s'}^\beta; (\rmg)\left[k_2^+,k_2\right]_{\eta}^c \right\rangle_0 \nonumber\\
&+&
A^{qg\gamma}\; g_sg_e\sum_{s's''}\sum_{\lambda\eta} \int \frac{dk_1^+}{2\pi}\frac{d^2k_1}{(2\pi)^2}\frac{dk_2^+}{2\pi}\frac{d^2k_2}{(2\pi)^2} \; t^c_{\alpha\beta}\nonumber\\
&&
\times
\Bigg\{ 
F^{(2)}_{(\rmq\rmp-\rmq\rmg)}\left[ (\rmp)[k_1^+,k_1]^\lambda , (\rmg)[k_2^+,k_2]^\eta , (\rmq)[p^+-k_1^+-k_2^+,-k_1-k_2]_{ss''} \right]
\nonumber\\
&&
\hspace{0.2cm}
+ \; 
F^{(2)}_{(\rmq\rmg-\rmq\rmp)}\left[ (\rmg)[k_2^+,k_2]^\eta , (\rmp)[k_1^+,k_1]^\lambda , (\rmq)[p^+-k_2^+-k_1^+,-k_2-k_1]_{ss''}\right]
\Bigg\}\nonumber\\
&&
\hspace{1.5cm}
\times
\left| (\rmq) \left[ p^+-k_1^+-k_2^+, -k_1-k_2\right]_{s''}^\beta , (\rmg) \left[k_2^+,-k_2 \right]^c_\eta , (\rmp) \left[k_1^+,k_1\right]^\lambda \right\rangle_0 \, .
\label{dwf1}
\eeq
Here, $A^q, A^{q\gamma}, A^{qg}$ and $A^{qg\gamma}$ are normalization constants whose explicit expression is not important for our purposes, $t^c$ are the generators of $SU(N_c)$ in the fundamental representation, and $g_e,g_s$ are the  QED and Yang-Mills coupling constants respectively.
 
 \begin{figure}[hbt]
\begin{center}
\vspace*{0.5cm}
\includegraphics[width=13cm]{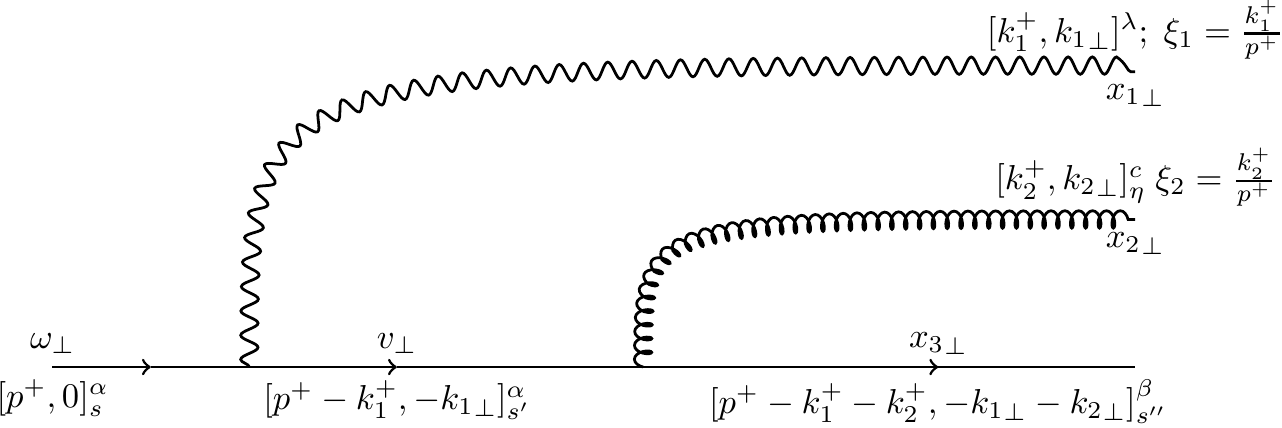}
\end{center}
\caption{The first component of the dressed quark wave function Eq.~\eqref{dwf1} where the photon is emitted before the gluon.}
\label{fig1}
\end{figure}
 
\begin{figure}[hbt]
\begin{center}
\vspace*{0.5cm}
\includegraphics[width=13cm]{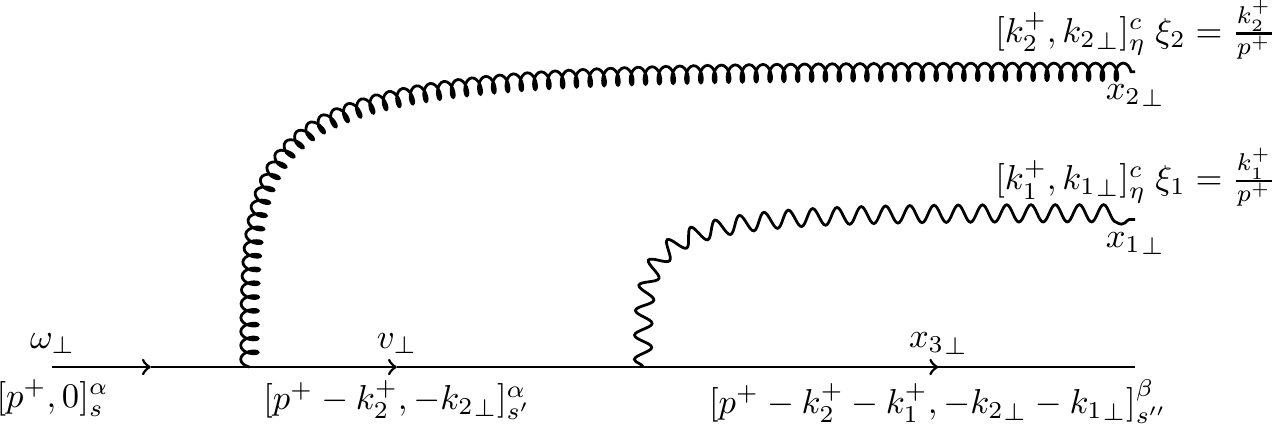}
\end{center}
\caption{The second component of the dressed quark wave function Eq.~ \eqref{dwf1} where the gluon is emitted before the photon.}
\label{fig2}
\end{figure}

The functions $F^{(1)}_{(\rmq\rmg)}$ and $F^{(1)}_{(\rmq\rmp)}$ are the well-known functions that define the momentum structure of the quark-gluon and quark-photon splitting amplitudes. The functions $F^{(2)}_{(\rmq\rmp-\rmq\rmg)}$ and $F^{(2)}_{(\rmq\rmg-\rmq\rmp)}$ define the momentum structure of two successive splittings. These two functions correspond to different orderings of photon and gluon emissions (see Figs. \ref{fig1} and \ref{fig2} where the kinematics and the notation for momenta and positions are also specified). The explicit expressions of these functions are well-known and can be obtained from e.g. \cite{Gelis:2002ki,Lublinsky:2016meo}. In the most general case (i.e. with non vanishing transverse momentum of the incoming quark $p$) the full momentum space expressions of these functions can be written as 

\beq
\label{F1}
F^{(1)}_{(\rmq\rmp)}\left[(\rmp)[k_1^+,k_1]^\lambda , (\rmq)[p^+-k_1^+,p-k_1]_{ss'}\right]=\bigg[\frac{-1}{\sqrt{2\xi_1p^+}}\phi^{\lambda\bar\lambda}_{ss'}(\xi_1)\bigg]\frac{\left(\xi_1p-k_1\right)^{\bar\lambda}}{\left(\xi_1p-k_1\right)^2}\ ,
\eeq
\beq
\label{F2}
&&
\hspace{-1.8cm}
F^{(2)}_{(\rmq\rmp-\rmq\rmg)}\left[ (\rmp)[k_1^+,k_1]^\lambda , (\rmg)[k_2^+,k_2]^\eta , (\rmq) [p^+-k_1^+-k_2^+,p-k_1-k_2]_{ss''}\right]\\
&&
= \sum_{s'} 
\bigg[\frac{1}{\sqrt{2\xi_1p^+}}\phi^{\lambda\bar\lambda}_{ss'}(\xi_1)\bigg]
\bigg[\frac{1}{\sqrt{2\xi_2p^+}}{\tilde\phi}^{\eta\bar\eta}_{s's''}(\xi_1,\xi_2)\bigg]\nonumber\\
&&
\hspace{0.3cm}
\times
\frac{(\xi_1p-k_1)^{\bar\lambda}}{(\xi_1p-k_1)^2}
\frac{[\xi_2(p-k_1)-\bar\xi_1k_2]^{\bar\eta}}{\xi_2(\xi_1p-k_1)^2+\xi_1(\xi_2p-k_2)^2-(\xi_2k_1-\xi_1k_2)^2}\ .\nonumber
\eeq
The expressions for $F^{(1)}_{(\rmq\rmg)}$  and $F^{(2)}_{(\rmq\rmg-\rmq\rmp)}$ are obtained by exchanging $\xi_1\leftrightarrow\xi_2$ and $k_1\leftrightarrow k_2$ in the expressions above. Here, we have defined
\beq
\label{phi}
\phi^{\lambda\bar\lambda}_{ss'}(\xi_1)=\Big[(2-\xi_1)\delta^{\lambda\bar\lambda}\delta_{ss'}-i\epsilon^{\lambda\bar\lambda}\sigma^3_{ss'} \xi_1\Big],
\eeq 
\beq  
\label{tildephi}
\tilde\phi^{\eta\bar\eta}_{s's''}(\xi_1,\xi_2)=\frac{\xi_1}{\bar\xi_1}\Big[(2-2\xi_1-\xi_2)\delta^{\eta\bar\eta}\delta_{s's''}-i\epsilon^{\eta\bar\eta}\sigma^3_{s's''}\xi_2\Big].
\eeq
Also, we have defined the $+$-momentum ratios as $k_1^+=\xi_1p^+$ and $k_2^+=\xi_2p^+$, $\bar \xi_1=1-\xi_1$, $\bar \xi_2=1-\xi_2$. The next step is to Fourier transform the dressed quark state and write it in the mixed $+$-momentum-transverse position space. Here, we only present the final result but the details of the calculation can be found in Appendix \ref{Section:Dressed_Quark_in_Mixed_State}. The dressed quark state in the mixed space reads
\beq
\label{Dressed_mixed_space}
&&
\hspace{-0.5cm}
\left| (\rmq) [p^+,0]^\alpha_s\right\rangle_D= 
\int_{\omega}\left| (\rmq) [p^+,\omega]_s^\alpha\right\rangle_0
\\ 
&+&
g_e\sum_{s'\lambda} \int \frac{dk_1^+}{2\pi} \int_{\omega, v, x_1} 
\bigg[\frac{(-i)}{\sqrt{2\xi_1p^+}}\,  \phi^{\lambda\bar\lambda}_{ss'}(\xi_1) \bigg]\, A^{\bar\lambda}(v-x_1) \, \delta^{(2)}\left[\omega-(\bar\xi_1v+\xi_1x_1)\right]
\nonumber\\
&&
\hspace{7cm}
\times
\left| (\rmq) \left[p^+-k_1^+,v\right]_{s'}^\alpha ; (\rmp) [k_1^+,{x_1}]_\lambda \right\rangle_0 \nonumber\\
&+&
g_s\sum_{s'\eta} \int \frac{dk_2^+}{2\pi}\int_{\omega, v, {x_2}} \, t^c_{\alpha\beta}\, 
\bigg[\frac{(-i)}{\sqrt{2\xi_2p^+}}\, \phi^{\eta\bar\eta}_{ss'}(\xi_2)\bigg] \, \bar{A}^{\bar\eta}(v-x_2) \, \delta^{(2)}\left[\omega-(\bar\xi_2v+\xi_2x_2)\right]
\nonumber\\
&&
\hspace{7cm}
\times
\left| (\rmq)\left[p^+-k_2^+,v\right]_{s'}^\beta; (\rmg)\left[k_2^+,{x_2}\right]_{\eta}^c \right\rangle_0 
\nonumber\\
&+&g_sg_e\sum_{s's''}\sum_{\lambda\eta} \int \frac{dk_1^+}{2\pi}\frac{dk_2^+}{2\pi}t^c_{\alpha\beta}
\int_{\omega,\,  v, \,x_1, \, x_2, \, x_3}\nonumber\\
&&
\times
\Bigg\{
\delta^{(2)}\left[v- \left\{\left(1-\frac{\xi_2}{\bar\xi_1}\right)x_3+\frac{\xi_2}{\bar\xi_1}x_2\right\}\right]
\delta^{(2)}\left[ \omega-\left(\xi_1x_1+\bar\xi_1v\right)\right]
\nonumber\\
&&
\hspace{0.7cm}
\times
\bigg[\frac{(-i)}{\sqrt{2\xi_1p^+}}\phi^{\lambda\bar\lambda}_{ss'}(\xi_1)\bigg]
\bigg[\frac{(-i)}{\sqrt{2\xi_2p^+}}\phi^{\eta\bar\eta}_{s's''}\left(\frac{\xi_2}{\bar\xi_1}\right)\bigg]
A^{\bar\eta}(x_3-x_2)\bar{\cal A}^{\bar\lambda}_{\xi_2/\bar\xi_1}(v-x_1)
\nonumber\\
&&
\hspace{0.5cm}
+\, 
\delta^{(2)}\left[v- \left\{\left(1-\frac{\xi_1}{\bar\xi_2}\right)x_3+\frac{\xi_1}{\bar\xi_2}x_1\right\}\right]
\delta^{(2)}\left[ \omega-\left(\xi_2x_2+\bar\xi_2v\right)\right]
\nonumber\\
&&
\hspace{0.7cm}
\times
\bigg[\frac{(-i)}{\sqrt{2\xi_2p^+}}\phi^{\eta\bar\eta}_{ss'}(\xi_2)\bigg]
\bigg[\frac{(-i)}{\sqrt{2\xi_1p^+}}\phi^{\lambda\bar\lambda}_{s's''}\left(\frac{\xi_1}{\bar\xi_2}\right)\bigg]
A^{\bar\lambda}(x_3-x_1)\bar{\cal A}^{\bar\eta}_{\xi_1/\bar\xi_2}(v-x_2)\Bigg\}
\nonumber\\
&&
\hspace{3.5cm}
\times
\left| (\rmq)\left[ p^+-k_1^+-k_2^+,{x_3}\right]_{s''}^\beta; (\rmg)\left[k_2^+,{x_2}\right]^c_\eta; (\rmp)\left[k_1^+,{x_1}\right]^\lambda \right\rangle_0\nonumber\,.
\eeq
Hereafter, integrals in transverse coordinate space are denoted as $\int_{\omega}\equiv \int d^2\omega$  
and  we have introduced $A^{\bar\lambda}(v-{x_1})$ as the electro-magnetic Weizs\''acker-Williams field in the quark-photon splitting function which is defined as
\beq
\label{Ajxi}
A^{\bar\lambda}(v-{x_1})= -\frac{1}{2\pi}\frac{(v-{x_1})^{\bar\lambda}}{(v-{x_1})^2}\; .
\eeq
We have also introduced several  variations of the Weizs\''acker-Williams field (denoted by $\bar A$ and $\bar {\cal A}$ in eq.(\ref{Dressed_mixed_space}), which take into account the Ioffe time constraint on the lifetime of fast fluctuations in the quark wave function  \cite{Gribov:1965hf, Ioffe:1969kf}. In the context of a similar hybrid calculation the Ioffe time constraint 
 was first introduced in \cite{nlohybridpart2}. The explicit expressions for various Ioffe time restricted Weizs\''acker-Williams fields can be found in Appendix A. For simplicity we will neglect this kinematic constraint in explicit calculations in the following.
If the Ioffe time restriction is neglected, then
\beq
\label{cal_A_without_Ioffe1}
\bar{\cal A}^{\bar\lambda}_{\xi_2/\bar\xi_1}(v-x_1)\rightarrow -\frac{1}{2\pi}\frac{\xi_1(v-x_1)^{\bar\lambda}}{\xi_1(v-x_1)^2+\frac{\xi_2}{\bar\xi_1}\left(1-\frac{\xi_2}{\bar\xi_1}\right)(x_3-x_2)^2}\equiv { \cal A}^{\bar\lambda}_{\xi_2/\bar\xi_1}(v-x_1).
\eeq

The initial dressed quark state (with vanishing transverse momentum) eikonally scatters through the target. Each bare state component of the dressed quark state given in Eq.\eqref{Dressed_mixed_space} rotates in the color space by picking up a fundamental or an adjoint $S$-matrix at the transverse position of the quark or the gluon that are defined in terms of the color fields of the target as
\beq
S(z)={\cal P}\, e^{ig\int dz^+ \tau^a A^-_a(z^+,z)},
\eeq
with $\tau^a$ being the generators of $SU(N_c)$ in the corresponding representation. The details of the calculation of the outgoing wave function in terms of dressed components are presented in  Appendix \ref{Section:Outgoing_WaveFunction}. 

The final result can be written  as
\beq
\label{out_full}
&&
\left| (\rmq)[p^+,0]_s^{\alpha}\right\rangle_{\rm out}= \int_\omega S_F^{\alpha\beta}(\omega) \left|(\rmq)[p^+,\omega]_s^{\beta}\right\rangle_D \\
&&
+g_e\sum_{s'\lambda}\int\frac{dk_1^+}{2\pi}\int_{\omega v x_1}
\left[S_F^{\alpha\beta}(v)-S_F^{\alpha\beta}(\omega)\right]
\bigg[\frac{(-i)}{\sqrt{2\xi_1p^+}}\phi^{\lambda\bar\lambda}_{ss'}(\xi_1)\bigg]
A^{\bar\lambda}(v-x_1)\nonumber\\
&&
\hspace{5cm}
\times
\delta^{(2)}\left[\omega-(\bar\xi_1v+\xi_1x_1)\right]
\left| (\rmq)[p^+-k_1^+,v]^\beta_{s}; (\rmp)[k_1^+,x_1]^\lambda\right\rangle_D\nonumber\\
&&
+g_s\sum_{s'\eta}\int\frac{dk_2^+}{2\pi}\int_{\omega v x_2}
\left[ t^c_{\alpha\beta}S_F^{\beta\sigma}(v)S^{cd}_A(x_2) - S_F^{\alpha\beta}(\omega) t^d_{\beta\sigma}\right] 
\bigg[\frac{(-i)}{\sqrt{2\xi_2p^+}} \phi^{\eta\bar\eta}_{ss'}(\xi_2)\bigg]
\bar{A}^{\bar\eta}_{\xi_2}(v-x_2) \nonumber\\
&&
\hspace{5cm}
\times
\delta^{(2)}\left[ \omega-(\bar\xi_2v+\xi_2x_2)\right]
\left| (\rmq)[p^+-k_2^+,v]^\sigma_{s'}; (\rmg)[k_2^+,x_2]^c_{\eta}\right\rangle_D\nonumber\\
&&
+g_sg_e \sum_{s's''}\sum_{\lambda\eta}\int \frac{dk_1^+}{2\pi}\frac{dk_2^+}{2\pi} \int_{wvx_1x_2x_3} 
\delta^{(2)}\Big[\omega-(\xi_1x_1+\bar\xi_1v)\Big]
\nonumber\\
&&
\times
\delta^{(2)}\bigg[v-\left\{\left(1-\frac{\xi_2}{\bar\xi_1}\right)x_3+\frac{\xi_2}{\bar\xi_1}x_2\right\}\bigg]
\bigg[\frac{(-i)}{\sqrt{2\xi_1p^+}} \phi^{\lambda\bar\lambda}_{ss'}(\xi_1)\bigg]
\bigg[\frac{(-i)}{\sqrt{2\xi_2p^+}} \phi^{\eta\bar\eta}_{s's''}\left(\frac{\xi_2}{\bar\xi_1}\right)\bigg]
\nonumber\\
&&
\times
\bigg\{
\left[ t^c_{\alpha\beta}S_F^{\beta\sigma}(x_3)S_A^{cd}(x_2) -S_F^{\alpha\beta}(\omega)t^d_{\beta\sigma}\right]
A^{\bar\eta}(x_3-x_2)
\bar{\cal A}^{\bar\lambda}_{\xi_2/\bar\xi_1}(v-x_1)\nonumber\\
&&
\hspace{2.5cm}
-
\left[S_F^{\alpha\beta}(v)-S_F^{\alpha\beta}(\omega)\right]t^d_{\beta\sigma} \, 
{\bar A}^{\bar\eta}_{\xi_2/\bar\xi_1}(x_3-x_2)
A^{\bar\lambda}(v-x_1)\bigg\}
\nonumber\\
&&
\hspace{5cm}
\times
\left| (\rmq)[p^+-k_1^+-k_2^+,x_3]^{\sigma}_{s''}, (\rmg)[k_2^+,x_2]^d_{\eta}, (\rmp) [k_1^+,x_1]^{\lambda}\right\rangle_D
\nonumber\\
&&
+g_sg_e \sum_{s's''}\sum_{\lambda\eta}\int \frac{dk_1^+}{2\pi}\frac{dk_2^+}{2\pi} \int_{wvx_1x_2x_3} 
\delta^{(2)}\Big[\omega-(\xi_2x_2+\bar\xi_2v)\Big]
\nonumber\\
&&
\times
\delta^{(2)}\bigg[v-\left\{\left(1-\frac{\xi_1}{\bar\xi_2}\right)x_3+\frac{\xi_1}{\bar\xi_2}x_1\right\}\bigg]
\bigg[\frac{(-i)}{\sqrt{2\xi_2p^+}} \phi^{\eta\bar\eta}_{ss'}(\xi_2)\bigg]
\bigg[\frac{(-i)}{\sqrt{2\xi_1p^+}} \phi^{\lambda\bar\lambda}_{s's''}\left(\frac{\xi_1}{\bar\xi_2}\right)\bigg]
A^{\bar\lambda}(x_3-x_1)
\nonumber\\
&&
\times
\bigg\{
\left[ t^c_{\alpha\beta}S_F^{\beta\sigma}(x_3)S_A^{cd}(x_2) -S_F^{\alpha\beta}(\omega)t^d_{\beta\sigma}\right]
\bar{\cal A}^{\bar\eta}_{\xi_1/\bar\xi_2}(v-x_2)\nonumber\\
&&
\hspace{2.5cm}
-
\left[ t^c_{\alpha\beta}S_F^{\beta\sigma}(v)S_A^{cd}(x_2) -S_F^{\alpha\beta}(\omega)t^d_{\beta\sigma}\right]
{\bar A}^{\bar\eta}_{\xi_1/\bar\xi_2}(v-x_2)\bigg\}
\nonumber\\
&&
\hspace{5cm}
\times
\left| (\rmq)[p^+-k_1^+-k_2^+,x_3]^{\sigma}_{s''}, (\rmg)[k_2^+,x_2]^d_{\eta}, (\rmp) [k_1^+,x_1]^{\lambda}\right\rangle_D\ .
\nonumber
\eeq

Let us make some  comments at this point. First, since we are interested in production of a photon and two jets, only the  dressed quark-photon-gluon component of the outgoing wave function  is relevant.
 Thus, for our purposes, we can neglect the dressed quark, dressed quark-photon and dressed quark-gluon components of the outgoing wave function.
 
Second, recall that we are interested in a very specific kinematics. We are considering the production of a soft photon with transverse momentum smaller than $Q_s$ and two hard jets with transverse momenta larger than $Q_s$. In this kinematics, the main production mechanism of the soft photon  is collinear radiation from the incoming quark with vanishing transverse momenta. The final hard momenta of the produced jets can arise from two different sources. The quark can split into a quark-gluon pair with small relative transverse momenta in the projectile wave function. In this case, the large transverse momentum of the outgoing jets comes entirely from  large momentum exchange between the target and each of the propagating  partons.
 In the second mechanism, the incoming quark splits into a quark-gluon pair with large relative transverse momenta already in the projectile wave function. In this case the leading contribution stems from small transverse momentum exchange during the interaction with the target. Such process leads  to almost back-to-back (in the transverse plane) jets - the situation that we analyze in Section \ref{Section:Back-to-back}.  
 It has been shown in \cite{Elastic_vs_Inelastic} that the second mechanism is sensitive to the saturation scale whereas the first one is not and provides a negligible contribution to the cross section. The same mechanism is also the dominant one in the situation of large momentum imbalance in the dilute target limit, as in this case only one of the partons needs to scatter in order to produce the requisite momentum imbalance. We  consider this situation in Section \ref{Section:Dilute target} in the  limit of a dilute target.
  
 Consequently, here we will concentrate on the contributions to the cross section which originate from hard quark-gluon splitting in the projectile wave function. Since the produced soft photon should be collinear to the original incoming quark, the photon must be emitted before the hard splitting (see Fig. \ref{fig1}). This allows us to neglect the contribution to the outgoing wave function when the photon is emitted after the hard splitting.

Finally, we would like to comment about the Ioffe time restriction. As it is discussed in detail in \cite{nlohybridpart2}, it is crucial in any NLO calculation since it directly affects the factorization scheme when one wants to account for evolution. We have derived the outgoing wave function taking into account the Ioffe time restriction for the sake of the completeness, and also with the idea that it can be used without further work for the calculation of other processes such as NLO photon production. However, in the explicit calculations of the cross section in this paper we will not impose the Ioffe time restriction on the phase space integrals in the wave function.

All in all, we can write the relevant part of the outgoing wave function (denoting it with the overline)  as 
\beq
\label{outbar}
&&\overline{\left| (\rmq) [p^+,0]_s^\alpha \right\rangle}_{\rm out}= g_sg_e\sum_{s's''}\sum_{\lambda\eta} \int \frac{dk_1^+}{2\pi}\frac{dk_2^+}{2\pi}
\int_{\omega,v,{x_1},{x_2},{x_3}} \delta^{2}\left[\omega-(\bar\xi_1v+\xi_1x_1)\right] 
\nonumber\\
&&
\times
\delta^{(2)}\left[v-(\bar\xi_2x_3+\xi_2x_2)\right] 
\left[\frac{(-i)}{\sqrt{2\xi_1p^+}}\phi^{\lambda\bar\lambda}_{ss'}(\xi_1)\right]
\left[\frac{(-i)}{\sqrt{2\bar{\xi_1}\xi_2p^+}}\phi^{\eta\bar\eta}_{s's''}(\xi_2)\right]
A^{\bar\eta}(x_3-x_2)
\nonumber\\
&&
\times
\Bigg\{ 
\left[t^c_{\alpha\beta}S_F^{\beta\sigma}(x_3)S_A^{cd}(x_2)-S_F^{\alpha\beta}(\omega)t^d_{\beta\sigma}\right]
{\cal A}^{\bar\lambda}_{\xi_2}(v-x_1)
-\left[ S_F^{\alpha\beta}(v)-S_F^{\alpha\beta}(\omega) \right]t^d_{\beta\sigma}A^{\bar\lambda}(v-x_1)\Bigg\}
\nonumber\\
&&
\times
\left| (\rmq)\left[ p^+-k_1^+-k_2^+,{x_3}_\perp\right]_{s''}^\sigma; (\rmg)\left[k_2^+,{x_2}\right]^d_\eta; (\rmp)\left[k_1^+,{x_1}\right]^\lambda\right\rangle_D \; ,
\eeq
where $A^{\bar\eta}(x_3-x_2)$ and $A^{\bar\lambda}(v-x_1)$ are the ordinary Weizs\''acker-Williams fields in the quark-gluon and quark-photon splittings respectively. On the other hand, ${\cal A}^{\bar\lambda}_{\xi_2/\bar\xi_1}(v-x_1)$ is the field  that appears in two successive emissions of the photon and gluon, defined in Eq.\eqref{cal_A_without_Ioffe1}.

When writing the relevant part of the outgoing wave function, Eq. \eqref{outbar}, we rescaled the $+$-momentum ratio; $\frac{\xi_2}{\bar\xi_1}\to\xi_2$, for convenience. This corresponds to defining the $+$-momentum ratio of the produced gluon with respect to the intermediate quark, rather than with respect to the initial quark. After this rescaling the $+$-momentum fractions carried by the photon and the gluon are defined as 
\beq
\frac{k_1^+}{p^+}=\xi_1 \; , \hspace{1cm} \frac{k_2^+}{p^+-k_1^+}=\xi_2\; .
\eeq

%
\subsection{Production cross section}
The production cross section can be written as a convolution of the quark distribution function inside the proton, $f^q_{\mu^2}(x_p)$, and the partonic level cross section as

\beq
\label{full_X_section}
\frac{d\sigma^{pA\to q\gamma g+X}}{d^3\q_1d^3\q_2d^3\q_3}= \int dx_p \, f^q_{\mu^2}(x_p)\, \frac{d\sigma^{qA\to q\gamma g+X}}{d^3\q_1d^3\q_2d^3\q_3}\,,
\eeq
where  $\mu^2$ is the factorization scale and $x_p$ is the $+$-momentum fraction carried by the incoming quark. Here, we introduced a short hand notation for the three-momenta, $\q_i\equiv(q_i^+,q_{i})$. The momenta $\q_1$, $\q_2$ and $\q_3$ are the three-momenta of the produced photon, gluon and quark respectively.  On partonic level, the production cross section of a photon and two jets is formally defined as the expectation value of the "number operator" in the outgoing wave function derived in the previous section:

\beq
&&
\hspace{-2.8cm}
(2\pi)^9\frac{d\sigma^{qA\to q\gamma g+X}}{d^3\q_1d^3\q_2d^3\q_3} (2\pi)\delta(p^+-q_1^+-q_2^+-q_3^+)=\nonumber\\
&&
\hspace{0.8cm}
= \frac{1}{2N_c}\sum_{s,\alpha}   {}_{\rm out}\overline{\left\langle ({\rmq})[{p}^+,0]_{s}^{\alpha} \right|} O(\q_1,\q_2,\q_3) \overline{\left | (\rmq)[p^+,0]_s^{\alpha}\right\rangle}_{\rm out}\,,
\eeq 
where the normalization factor $1/2N_c$ comes from averaging over the color and spin indexes in the incoming wave function in the amplitude and complex conjugate amplitude. The number operator is defined as 
\beq
\hspace{-0.8cm}
O(\q_1,\q_2,\q_3)=
\gamma^\dagger_{\lambda}(\q_1)\gamma_{\lambda}(\q_1)
a^{\dagger b}_i(\q_2)a^b_i(\q_2)d^{\dagger\beta}_t(\q_3)
d^{\beta}_t(\q_3)\, .
\eeq
Here, $\gamma^\dagger_\lambda(\q_1)$ is the creation operator of a dressed photon with three-momentum $\q_1$ and polarization $\lambda$, $a^{\dagger b}_i(\q_2)$ is the creation operator of a dressed gluon with three-momentum $\q_2$, color $b$ and polarization index $i$, and $d^{\dagger\beta}_t (\q_3)$ is the creation operator for a quark with three-momentum $\q_3$, color $\beta$ and spin $t$. When written in the mixed space, the expectation value of the number operator reads
\beq
&&
\hspace{-1.5cm}
\Big\langle O (\q_1,\q_2,\q_3) \Big\rangle_{\rm out}= \int_{y_1z_1,y_2z_2,y_3z_3}e^{iq_1\cdot (y_1-z_1)+iq_2\cdot (y_2-z_2)+iq_3\cdot (y_3-z_3)}\nonumber\\
&&
\hspace{1.5cm}
\times \; \; 
 {}_{\rm out}\overline{\left\langle ({\rmq})[{p}^+,0]_{s}^{\alpha} \right|} \gamma^\dagger_{\lambda}(q_1^+,y_{1})\gamma_{\lambda}(q_1^+,z_{1})
 a^{\dagger b}_i(q_2^+,y_{2})\\
 &&
\hspace{4cm}
\times\; \; \; \; 
a^b_i(q_2^+,z_{2})
d^{\dagger\beta}_t(q_3^+,y_{3})
d^{\beta}_t(q_3^+,z_{3})
\overline{\left | (\rmq)[p^+,0]_s^{\alpha}\right\rangle}_{\rm out}\,. \nonumber
\eeq
The action of the creation and annihilation operators on the dressed one particle states and on the Fock vacuum is defined in the usual way. For example, for gluons we have
\beq
\label{g_cre}
a^b_i(q_2^+,z_2)\left| (\rmg)[k_2^+,x_2]_\eta^d\right\rangle_D&=& 2\pi \, \delta^{bd} \, \delta_{\eta i} \, 
\delta(k_2^+-q_2^+) \, \delta^{(2)}(x_2-z_2)|0\rangle, \\
\label{ani}
a^{\dagger b}_i(q_2^+,y_2)|0\rangle&=&\left| (\rmg)[q^+_2,y_2]_i^b\right\rangle_D \; .
\eeq
Analogous equations hold for the action of quark and photon creation and annihilation operators.  
Using these equations, one can calculate the partonic level production cross section. After some algebra the result reads
\beq
\label{Partonic_X_section_1}
&&
(2\pi)^9\frac{d\sigma^{qA\to q\gamma g+X}}{d^3\q_1d^3\q_2d^3\q_3} =  \frac{1}{2N_c}g_s^2g_e^2 \; 
\sum_{s's''} \sum_{\lambda\eta} \, (2\pi) \, \delta(p^+-q_1^+-q_2^+-q_3^+)\\
&&
\times
 \int_{y_1z_1,y_2z_2,y_3z_3,\omega v,\omega'  v'}
e^{iq_1\cdot(y_1-z_1)+iq_2\cdot(y_2-z_2)+iq_3\cdot(y_3-z_3)} \, 
\delta^{(2)}[\omega-(\bar\xi_1v+\xi_1z_1)] \, 
\nonumber\\
&&
\times \, 
\delta^{(2)}[\omega'-(\bar\xi_1v'+\xi_1y_1)]\, 
\delta^{(2)}[v-(\bar\xi_2z_3+\xi_2z_2)]\, 
\delta^{(2)}[v'-(\bar\xi_2y_3+\xi_2y_2)]\, 
\nonumber\\
&&
\times \,
A^{\bar\eta}(z_3-z_2) \,
A^{\bar\eta'}(y_3-y_2)
\left\{ \frac{ \phi^{\lambda\bar\lambda}_{ss'}(\xi_1)}{\sqrt{2\xi_1p^+}} \, \frac{\phi^{*\lambda\bar\lambda'}_{s\bar{s}'}(\xi_1)}{\sqrt{2\xi_1p^+}}   \right\} 
\left\{ \frac{\phi^{\eta\bar\eta}_{s's''}(\xi_2)}{\sqrt{2\bar\xi_1\xi_2p^+}} \, \frac{\phi^{*\eta\bar\eta'}_{\bar{s}'s''}(\xi_2)}{\sqrt{2\bar\xi_1\xi_2p^+}} \right\}
\nonumber\\
&&
\times \, 
\Bigg\langle
\bigg\{ \!\!
\left[  S^{\dagger\bar{c}'d}_A(y_2)S^\dagger_F(y_3)t^{\bar{c}'}   -   t^dS^\dagger_F(\omega')   \right]_{\sigma\alpha}
\!\! \!\! \ {\cal A}_{\xi_2}^{\bar\lambda'}(v'-y_1)
 - \left[ t^d( S^\dagger_F(v')-S^\dagger_F(\omega'))\right]_{\sigma\alpha} 
\!\!\!\!
 A^{\bar\lambda'}(v'-y_1)
\bigg\}
\nonumber\\
&&
\times
\,
\bigg\{
\left[ t^cS_F(z_3)S^{cd}_A(z_2)-S_F(\omega)t^d\right]_{\alpha\sigma} 
\!\!\!\! {\cal A}^{\bar\lambda}_{\xi_2}(v-z_1)
-\left[ \left(S_F(v)-S_F(\omega)\right)t^d\right]_{\alpha\sigma}
\!\!\!\!
A^{\bar\lambda}(v-z_1)
\bigg\} \Bigg\rangle_T\ ,
\nonumber
\eeq
where $\left\langle \cdots \right\rangle_T$ denotes averaging over the ensemble of the target fields that has to be performed to obtain the final result (sometimes this average will not be indicated in the intermediate expressions and will be reinstated in the final formulae).
Using the explicit expression for the splitting amplitudes that was defined in Eq. \eqref{phi}, the square of the splitting amplitudes can be calculated in a straightforward manner:
\beq
\phi^{\lambda\bar\lambda}_{ss'}(\xi_1)\phi^{*\lambda\bar\lambda'}_{s\bar{s}'}(\xi_1)&=&
\delta_{s'\bar{s}'}\delta^{\bar\lambda\bar{\lambda}'}2(1+\bar\xi_1^2),\\
\phi^{\eta\bar\eta}_{s's''}(\xi_2)\phi^{*\eta\bar\eta'}_{\bar{s}'s''}(\xi_2)&=&
\delta_{s'\bar{s}'}\delta^{\bar\eta\bar{\eta}'}2(1+\bar\xi_2^2).
\eeq
The parton level production cross section then reads
\beq
\label{Partonic_X_section_2}
&&
(2\pi)^9\frac{d\sigma^{qA\to q\gamma g+X}}{d^3\q_1d^3\q_2d^3\q_3} =  \frac{1}{N_c}g_s^2g_e^2 \;
(2\pi)\delta(p^+-q_1^+-q_2^+-q_3^+)  \, 
\frac{1}{\xi_1p^+} \, 
\frac{1}{\bar\xi_1\xi_2p^+}
\\
&&
\times
 \int_{y_1z_1,y_2z_2,y_3z_3,\omega v,\omega'  v'}
e^{iq_1\cdot(y_1-z_1)+iq_2\cdot(y_2-z_2)+iq_3\cdot(y_3-z_3)} \nonumber\\
&&
\times
\delta^{(2)}[\omega-(\bar\xi_1v+\xi_1z_1)] \, 
\delta^{(2)}[\omega'-(\bar\xi_1v'+\xi_1y_1)]\,
\delta^{(2)}[v-(\bar\xi_2z_3+\xi_2z_2)]\, 
\nonumber\\
&&
\times \, 
\delta^{(2)}[v'-(\bar\xi_2y_3+\xi_2y_2)]\, 
A^{\eta}(z_3-z_2) \,
A^{\eta}(y_3-y_2) \, 
(1+\bar\xi_1^2) \, (1+\bar\xi_2^2)
\nonumber\\
&&
\times \, 
\bigg\{{\cal A}_{\xi_2}^{\lambda}(v'-y_1) {\cal A}_{\xi_2}^{\lambda}(v-z_1) 
\nonumber\\
&&
\hspace{2cm}
\times
\tr\Big[
\Big( S^{\dagger\bar{c}'d}_A(y_2)S^\dagger_F(y_3)t^{\bar{c}'}-t^dS^\dagger_F(\omega')\Big)
\Big( t^cS_F(z_3)S_A^{cd}(z_2)-S_F(\omega)t^d\Big)
\Big]
\nonumber\\
&&
\hspace{0.5cm}
+A^\lambda(v'-y_1) A^\lambda(v-z_1)
\tr\Big[t^d\Big(S^\dagger_F(v')-S^\dagger_F(\omega')\Big)\Big(S_F(v)-S_F(\omega)\Big)t^d\Big]\nonumber\\
&&
\hspace{0.5cm}
-{\cal A}_{\xi_2}^\lambda(v'-y_1) A^\lambda(v-z_1) \tr\Big[ \Big(  S^{\dagger\bar{c}'d}_A(y_2)S^\dagger_F(y_3)t^{\bar{c}'}-t^dS^\dagger_F(\omega')\Big) \Big(S_F(v)-S_F(\omega)\Big)t^d\Big]\nonumber\\
&&
\hspace{0.5cm}
-A^\lambda(v'-y_1)\,{\cal A}_{\xi_2}^\lambda(v-z_1)
\tr\Big[ t^d\Big(S^\dagger_F(v')-S^\dagger_F(\omega')\Big) \Big( t^cS_F(z_3)S_A^{cd}(z_2)-S_F(\omega)t^d\Big) \Big] 
\bigg\}.\nonumber
\eeq
Eq. \eqref{Partonic_X_section_2} can be further simplified. Apart from performing the straightforward $\delta$-function integrals, one can also simplify the Wilson line structure by using the Fierz identity
\beq
\label{Fierz}
t^a_{\alpha\beta}t^a_{\sigma\lambda}=\frac{1}{2}\left[\delta_{\alpha\lambda}\delta_{\beta\sigma}-\frac{1}{N_c}\delta_{\alpha\beta}\delta_{\sigma\lambda}\right]
\eeq
and the identity that relates adjoint and fundamental representations of a unitary matrix,
\beq
\label{Adj_to_Fund}
S^{ab}_A(x)=2\tr\left[t^aS_F(x)t^bS_F^\dagger(x)\right] \, .
\eeq 
After some color algebra, the cross section reads
\beq
\label{Partonic_X_section_final}
&&
(2\pi)^9\frac{d\sigma^{qA\to q\gamma g+X}}{d^3\q_1d^3\q_2d^3\q_3} =  \frac{1}{N_c}g_s^2g_e^2 \;
(2\pi)\delta(p^+-q_1^+-q_2^+-q_3^+)  \, 
\frac{1}{\xi_1p^+} \, 
\frac{1}{\bar\xi_1\xi_2p^+}
\\
&&
\times
 \int_{y_1z_1,y_2z_2,y_3z_3,\omega v,\omega'  v'}
e^{iq_1\cdot(y_1-z_1)+iq_2\cdot(y_2-z_2)+iq_3\cdot(y_3-z_3)} \nonumber\\
&&
\times
\delta^{(2)}[\omega-(\bar\xi_1v+\xi_1z_1)] \, 
\delta^{(2)}[\omega'-(\bar\xi_1v'+\xi_1y_1)]\,
\delta^{(2)}[v-(\bar\xi_2z_3+\xi_2z_2)]\, 
\nonumber\\
&&
\times \, 
\delta^{(2)}[v'-(\bar\xi_2y_3+\xi_2y_2)]\, 
A^{\eta}(z_3-z_2) \,
A^{\eta}(y_3-y_2) \, 
(1+\bar\xi_1^2) \, (1+\bar\xi_2^2)
\nonumber\\
&&
\times \, 
\bigg\{{\cal A}_{\xi_2}^{\lambda}(v'-y_1) {\cal A}_{\xi_2}^{\lambda}(v-z_1) 
\bigg[ 
\frac{N_c^2}{2}
\bigg( s(z_2,y_2)Q(y_2,z_2,z_3,y_3)+s(\omega,\omega')
\nonumber\\
&&
\hspace{0.5cm}
-s(\omega,y_2)s(y_2,y_3)-s(z_2,\omega')s(z_3,z_2)\bigg)
+
\frac{1}{2} \bigg( s(z_3,\omega')+s(\omega,y_3)-s(z_3,y_3)-s(\omega,\omega')\bigg)\bigg]\nonumber\\
&&
\hspace{0.5cm}
+ \, 
A^\lambda(v'-y_1)A^\lambda(v-z_1)\bigg[
\frac{N_c^2-1}{2}\bigg( s(v,v')+s(\omega,\omega')-s(v,\omega')-s(\omega,v')  \bigg)\bigg]\nonumber\\
&&
\hspace{0.5cm}
- \, 
{\cal A}_{\xi_2}^\lambda(v'-y_1)A^\lambda(v-z_1)\bigg[ 
\frac{N_c^2}{2}
\bigg( s(y_2,y_3)\Big[s(v,y_2)-s(\omega,y_2)\Big] -s(v,\omega')+s(\omega,\omega')\bigg)\nonumber\\
&&
\hspace{6.5cm}
+\frac{1}{2}\bigg( s(\omega,y_3)+s(v,\omega')-s(v,y_3)-s(\omega,\omega')\bigg)\bigg]\nonumber\\
&&
\hspace{0.5cm}
- \, 
A^\lambda(v'-y_1){\cal A}_{\xi_2}^\lambda(v-z_1)\bigg[ \frac{N_c^2}{2}\bigg( 
s(z_3,z_2)\Big[s(z_2,v')-s(z_2,\omega')\Big] -s(\omega,v')+s(\omega,\omega')\bigg)\nonumber\\
&&
\hspace{6.5cm}
+\, 
\frac{1}{2}\bigg( s(z_3,\omega')+s(\omega,v')-s(z_3,v')-s(\omega,\omega')\bigg)\bigg]\bigg\},\nonumber
\eeq
where we have defined the fundamental quadrupole and dipole operators as
\beq
Q(x,y,u,v)=\frac{1}{N_c}\tr\left[ S_F(x)S_F^\dagger(y)S_F(u)S_F^\dagger(v)\right] ,\;  \; 
s(x,y)=\frac{1}{N_c}\tr\left[ S_F(x)S_F^\dagger(y)\right] .
\eeq
The operators are defined such that $Q=1$ and $s=1$ for  vanishing background field. In this limit  the total cross section vanishes as it should.

The integrations over $\omega, v,\bar{\omega}$ and $\bar{v}$  can be trivially performed, but we choose to leave the expression in the above form  as it is more compact. To obtain the p-A cross section one should convolute the partonic level result, Eq. \eqref{Partonic_X_section_final}, with the quark distribution function inside the proton as mentioned earlier (see Eq. \eqref{full_X_section}).

\section{The back-to-back correlation limit}
\label{Section:Back-to-back}
For production of jets with transverse momenta $|q_2|$ and $|q_3|$ much larger than the saturation momentum of the target, $|q_2|,|q_3|\gg Q_s$, our expressions can be simplified further. As discussed earlier, the origin of the hard momenta of the produced jets is the large relative transverse momenta of the split quark-gluon pair in the wave function. When the transverse momentum transfer between the target and the quark-gluon pair during the interaction is small, the final jets will propagate almost back-to-back in the transverse plane. The small transverse momentum imbalance of the jets, $|{q}_{2}+{q}_{3}|$, is then sensitive to the transverse momenta of the gluons in the target which are on the order of the saturation scale, i.e., $|{q}_{2}+{q}_{3}|\sim Q_s$. This corresponds to a large relative momentum of the produced jets, $|q_3-q_2|\gg Q_s$. Therefore, we are interested in the kinematics: $|q_2|$, $|q_3|$, $|{q}_{2}-{q}_{3}|$ $\gg$ $|q_1|$, $|{q}_{2}+{q}_{3}|$ $\sim$ $Q_s$. In this situation the transverse size of the produced quark-gluon pair in the coordinate space is small. This allows us to utilize a small dipole approximation and expand our final result in powers of the dipole sizes. 

We start with the production cross section, Eq. \eqref{Partonic_X_section_final}, derived in the previous section and perform the following change of variables:
\beq
r&=&z_3-z_2 \, , \hspace{2cm} b=\frac{1}{2}(z_2+z_3)\ , \\
\bar{r}&=&y_3-y_2  \, , \hspace{2cm} \bar{b}=\frac{1}{2}(y_2+y_3)\ .
\eeq
Here, $r$ and $\bar r$ correspond to the transverse sizes of the produced quark-gluon pair in the amplitude and complex conjugate amplitude respectively. It is also convenient to define the relative transverse position of the produced soft photon by shifting the variables $y_1$ and $z_1$ in the following way:
\beq
\gamma&=&z_1-b-(1-2\xi_2)\frac{r}{2}\ ,\\
\bar\gamma&=&y_1-\bar b-(1-2\xi_2)\frac{\bar r}{2}\ .
\eeq
After performing these changes of variables, the parton level production cross section reads
\beq
\label{X_section_in_r_b}
&&
(2\pi)^9\frac{d\sigma^{qA\to q\gamma g+X}}{d^3\q_1d^3\q_2d^3\q_3}=\frac{1}{N_c}g_s^2g_e^2\,(2\pi)\,\delta(p^+-q_1^+-q_2^+-q_3^+)\,  \frac{1}{\xi_1p^+} \, \frac{1}{\bar\xi_1\xi_2p^+}\\
&&
\times
\int_{\gamma \bar\gamma b\bar{b}r\bar{r}\omega\omega'vv'} \!\!\!\!\!\!\!\!\!\!\!
{\rm exp}\left\{iq_1\cdot(\bar\gamma-\gamma)+i(q_1+q_2+q_3)\cdot(\bar b-b)+\frac{i}{2}\left[(1-2\xi_2)q_1-q_2+q_3\right]\cdot(\bar r-r)\right\}
\nonumber\\
&&
\times
\delta^{(2)}\left[\omega-\left( b+\xi_1\gamma+(1-2\xi_2)\frac{r}{2}\right) \right] 
\delta^{(2)} \left[ \omega'-\left( \bar{b}+\xi_1\bar{\gamma}+(1-2\xi_2)\frac{\bar{r}}{2}\right)\right]\nonumber\\
&&
\times
\delta^{(2)}\left[ v-\left( b+(1-2\xi_2)\frac{r}{2}\right)\right]
\delta^{(2)}\left[ v'-\left(\bar{b}+(1-2\xi_2)\frac{\bar{r}}{2}\right)\right]
(1+\bar\xi_1^2)(1+\bar\xi_2^2)
A^\eta(r) A^\eta(\bar{r})
\nonumber\\
&&
\times 
\bigg\{ {\cal A}_{\xi_2}^\lambda(-\bar\gamma) {\cal A}_{\xi_2}^\lambda(-\gamma)
\bigg[\frac{N_c^2}{2}\bigg( s\Big(b-\frac{r}{2},\bar{b}-\frac{\bar{r}}{2}\Big) Q\Big( \bar{b}-\frac{\bar r}{2}, b-\frac{r}{2}, b+\frac{r}{2}, \bar{b}+\frac{\bar r}{2}\Big)+s(\omega,\omega') \nonumber\\
&&
\hspace{4cm}
-s\Big(b-\frac{r}{2},\omega'\Big)s\Big( b+\frac{r}{2}, b-\frac{r}{2}\Big)-s\Big(\omega, \bar{b}-\frac{\bar r}{2}\Big)s\Big(\bar{b}-\frac{\bar r}{2}, \bar{b}+\frac{\bar r}{2}\Big)\bigg)
\nonumber\\
&&
\hspace{4cm}
+
\frac{1}{2}\bigg( s\Big(b+\frac{r}{2},\omega'\Big)+s\Big(\omega, \bar{b}+\frac{\bar r}{2}\Big)-s\Big(b+\frac{r}{2}, \bar{b}+\frac{\bar r}{2}\Big)-s(\omega,\omega')\bigg)\bigg] \nonumber\\
&&
+ A^{\lambda}(-\bar\gamma)A^\lambda(-\gamma)\bigg[\frac{N_c^2-1}{2}\bigg( s(v,v')+s(\omega,\omega')-s(v,\omega')-s(\omega,v')\bigg)\bigg]
\nonumber\\
&&
-
{\cal A}_{\xi_2}^\lambda(-\bar\gamma)A^\lambda(-\gamma)\bigg[\frac{N_c^2}{2}\bigg(
s\Big(\bar{b}-\frac{\bar r}{2}, \bar{b}+\frac{\bar r}{2}\Big) \Big[
s\Big(v, \bar{b}-\frac{\bar r}{2}\Big)-s\Big(\omega, \bar{b}-\frac{\bar r}{2}\Big) \Big]-s(v,\omega')+s(\omega,\omega')\bigg)
\nonumber\\
&&
\hspace{5cm}
+
\frac{1}{2}\bigg(s\Big(\omega, \bar{b}+\frac{\bar r}{2}\Big)+s(v,\omega')-s\Big(v, \bar{b}+\frac{\bar r}{2}\Big)-s(\omega,\omega')\bigg)\bigg]
\nonumber\\
&&
-A^\lambda(-\bar\gamma){\cal A}_{\xi_2}^\lambda(-\gamma)\bigg[ 
\frac{N_c^2}{2}\bigg( 
s\Big(b+\frac{r}{2},b-\frac{r}{2}\Big)\Big[s\Big(b-\frac{r}{2},v'\Big)-s\Big(b-\frac{r}{2},\omega'\Big)\Big]-s(\omega,v')+s(\omega,\omega')\bigg)
\nonumber\\
&&
\hspace{5cm}
+
\frac{1}{2}\bigg( 
s\Big(b+\frac{r}{2},\omega'\Big)+s(\omega,v')-s\Big(b+\frac{r}{2},v'\Big)-s(\omega,\omega')\bigg)\bigg]\bigg\}.\nonumber
\eeq
In this expression the conjugate momentum to $\bar\gamma-\gamma$ is $q_{1}$, the conjugate momentum to $\bar b-b$ is $ q_{1}+q_{2}+q_{3}$ and the conjugate momentum to $\bar r-r$ is $[(1-2\xi_2)q_{1}-q_{2}+q_{3}]/2$. In our kinematics $ |q_{1}+q_{2}+q_{3}| \approx |q_{2}+q_{3}|$ and  $|(1-2\xi_2)q_{1}-q_{2}+q_{3}|\approx |q_{3}-q_{2}|$. Therefore, in the back-to-back limit $ |\bar r|,|r| \ll |\bar\gamma|,|\gamma|$ and $ |\bar r|,|r|\ll |\bar b|,|b|$. Now, we can perform the small dipole approximation which amounts to  Taylor expanding the dipole and the quadrupole operators  as well as ${\cal A}_{\xi_2}^{\lambda}(-\bar\gamma)$ and ${\cal A}_{\xi_2}^{\lambda}(-\gamma)$ in powers of $r$ and $\bar r$. Here, we only present the final result for this expansion with the details given in Appendix \ref{Section:AppendixB}. 

The first non-vanishing term in the expansion is ${\cal O}(r\bar r)$ and the production cross section can be written as  
\beq
&&
(2\pi)^9\frac{d\sigma^{qA\to q\gamma g+X}}{d^3\underline{q}_1 d^3\underline{q}_2 d^3\underline{q}_3}=
\frac{1}{N_c} \, g_s^2g_e^2 \, (2\pi)\delta(p^+-q_1^+-q_2^+-q_3^+)\, \frac{1}{\xi_1p^+} \, \frac{1}{\bar\xi_1\xi_2p^+}
(1+\bar\xi_1^2) \, (1+\bar\xi_2^2)\nonumber\\
&&
\times \, 
\int_{r\bar{r}b\bar{b}\gamma\bar\gamma}
e^{iq_1\cdot(\bar\gamma-\gamma)+i(q_1+q_2+q_3)\cdot(\bar b-b) + \frac{i}{2}[(1-2\xi_2)q_1-q_2+q_3]\cdot(\bar r-r)}
A^\eta(r)A^\eta(\bar r) A^\lambda(\gamma)A^\lambda(\bar \gamma)\nonumber
\\
&&
\times \, 
\frac{N_c^2-1}{2}
r^i\bar r^j
\Bigg\{ 
\bigg[\xi_2^2-\frac{(1-2\xi_2)}{N_c^2-1}\bigg]
\bigg\langle
\frac{1}{N_c}\tr\Big(\partial^iS_F(b)\partial^jS_F^\dagger({\bar b})\Big)
\bigg\rangle_T
\nonumber\\
&&
\hspace{4cm}
-\frac{N_c^2}{N_c^2-1}
\bigg\langle\frac{1}{N_c}\tr\Big(\partial^iS_F(b)S^\dagger_F({\bar b})\partial^jS_F({\bar b})S_F^\dagger(b)\Big)s(b,\bar b)
\bigg\rangle_T
\Bigg\}
\ ,
\eeq
%
where we have reintroduced  the average over the target color field configuration that has to be performed to obtain the final result. 

Our next order of business is performing the integrations over $r$, $\bar r$, $\gamma$ and $\bar \gamma$. Let us first consider the integrations over $r$ and $\bar{r}$ which are factorized from the rest of the expression. After defining the conjugate momenta to the difference between the dipole sizes in the amplitude and complex conjugate amplitude, $(\bar r- r)$, as 
\beq
K_T\equiv \frac{1}{2}\left[ (1-2\xi_2)q_1-q_2+q_3 \right] \, , 
\eeq
and using the explicit expression for the modified Weizs\''acker-Williams field Eq. \eqref{Modified_A},
the integration over $r$ and $\bar r$ can be performed in a straightforward manner:
\beq
&&
\int_{r\bar r} e^{iK_T\cdot (\bar r-r)} \frac{r\cdot \bar r}{r^2\bar r^2} r^i\bar r^j
= (2\pi)^2\, \frac{\delta^{ij}}{K_T^4}\ .
\eeq
After integrating over $r$ and $\bar r$, the production cross section can be written as 
\beq
&&
\hspace{-0.4cm}
(2\pi)^9\frac{d\sigma^{qA\to q\gamma g+X}}{d^3\q_1d^3\q_2d^3\q_3}= 
(2\pi)\delta(p^+-q_1^+-q_2^+-q_3^+) \, 4C_F\, \alpha_s\alpha_e \, (2\pi)^2 \frac{1}{\xi_1p^+}\frac{1}{\xi_2\bar\xi_1p^+}
(1+\bar\xi_1^2)\, (1+\bar\xi_2^2)
\nonumber\\
&&
\times \, 
\frac{1}{K_T^4}\frac{1}{(2\pi)^2}
\int_{b\bar b, \gamma\bar\gamma}e^{iq_1\cdot(\bar\gamma-\gamma)+iP_T\cdot(\bar b-b)}
\frac{\gamma\cdot\bar\gamma}{\gamma^2\bar\gamma^2}
\Bigg\{
\bigg[ \xi_2^2-\frac{(1-2\xi_2)}{N_c^2-1}\bigg]
\bigg\langle \frac{1}{N_c} \tr\Big(\partial^iS_F(b)\partial^iS_F^\dagger({\bar b})\Big)
\bigg\rangle_T 
\nonumber\\
&&
\hspace{4.5cm}
-\frac{N_c^2}{N_c^2-1}\bigg\langle \frac{1}{N_c}\tr\Big(\partial^iS_F(b)S_F^\dagger({\bar b})\partial^iS_F({\bar b})S_F^\dagger(b)\Big)
s(b,\bar b)\bigg\rangle_T \Bigg\}
\ ,
\eeq
where we have defined the total transverse momentum of the produced particles  conjugate to $(\bar b-b)$ as
\beq
P_T\equiv q_1+q_2+q_3 \; .
\eeq
Note that the integrations over $\gamma$ and $\bar \gamma$  are factorized from the rest of the expression as well. We use the identity
\beq
\frac{k^i}{k^2}=\frac{1}{2\pi i}\int_z e^{ik\cdot z}\frac{z^i}{z^2}
\eeq
to integrate over  $\gamma$ and $\bar\gamma$,  to get
\beq
\label{finalwf}
&&
\hspace{-0.3cm}
(2\pi)^9\frac{d\sigma^{qA\to q\gamma g+X}}{d^3\q_1d^3\q_2d^3\q_3}= 
(2\pi)\delta(p^+-q_1^+-q_2^+-q_3^+) \,4 C_F\, \alpha_s\alpha_e \, (2\pi)^2 \frac{1}{\xi_1p^+}\frac{1}{\xi_2\bar\xi_1p^+}
(1+\bar\xi_1^2)
(1+\bar\xi_2^2)
\, \nonumber\\
&&
\hspace{-0.2cm}
\times \, 
\frac{1}{K_T^4}\frac{1}{q_1^2}
\int_{b\bar b}e^{iP_T\cdot(\bar b-b)}
\Bigg\{
\bigg[ \xi_2^2-\frac{(1-2\xi_2)}{N_c^2-1}\bigg]
\bigg\langle \frac{1}{N_c} \tr\Big(\partial^iS_F(b)\partial^iS_F^\dagger({\bar b})\Big)
\bigg\rangle_T 
\nonumber\\
&&
\hspace{4cm}
-\frac{N_c^2}{N_c^2-1}\bigg\langle \frac{1}{N_c}\tr\Big(\partial^iS_F(b)S_F^\dagger({\bar b})\partial^iS_F({\bar b})S_F^\dagger(b)\Big)
s(b,\bar b)\bigg\rangle_T \Bigg\} \, .
\eeq
To obtain the final result this parton level cross section  has to be convoluted with the parton density function, Eq. \eqref{full_X_section}. Using $x_p=p^+/ p_p^+$, the full production cross section gives 
\beq
\label{finalwfwithpdf}
&&
\hspace{-0.3cm}
(2\pi)^9\frac{d\sigma^{pA\to q\gamma g+X}}{d^3\q_1d^3\q_2d^3\q_3}=\!\! \int \! dx_p  \,x_p\, f^q_{\mu^2}(x_p)
\delta \!\left( \! x_p\!-\!\frac{q_1^++q_2^++q_3^+}{p_p^+} \right)
\!(2\pi)^3\,  4C_F\, \alpha_s\alpha_e \, \frac{1}{ p^+} \frac{1}{\xi_1 p^+}\frac{1}{\xi_2\bar\xi_1p^+}\, 
\nonumber\\
&&
\times
\,
(1+\bar\xi_1^2) \, (1+\bar\xi_2^2) 
\frac{1}{K_T^4}\frac{1}{q_1^2} \, 
\int_{b\bar b}e^{iP_T\cdot(\bar b-b)}
\Bigg\{
\bigg[ \xi_2^2-\frac{(1-2\xi_2)}{N_c^2-1}\bigg]
\bigg\langle \frac{1}{N_c} \tr\Big(\partial^iS_F(b)\partial^iS_F^\dagger({\bar b})\Big)
\bigg\rangle_T 
\nonumber\\
&&
\hspace{4cm}
-\frac{N_c^2}{N_c^2-1}\bigg\langle \frac{1}{N_c}\tr\Big(\partial^iS_F(b)S_F^\dagger({\bar b})\partial^iS_F({\bar b})S_F^\dagger(b)\Big)
s(b,\bar b)\bigg\rangle_T \Bigg\} \, .
\eeq
Eq. \eqref{finalwfwithpdf} is the final result for the fixed order production cross section of a soft photon plus two hard jets in the back-to-back correlation limit.
%

This expression has to be supplemented with the appropriate Sudakov factor \cite{Sudakov:1954sw} that resums the double logarithms coming from higher order emissions. This aspect has been extensively studied previously, see \cite{Mueller:2013wwa}, and we simply modify our expressions following the results of \cite{Mueller:2013wwa}. 
The resulting expression 
reads
\beq
&&
\hspace{-0.3cm}
(2\pi)^9\frac{d\sigma^{pA\to q\gamma g+X}}{d^3\q_1d^3\q_2d^3\q_3}=\!\! \int \! dx_p  \,x_p\, f^q_{\mu^2}(x_p)
\delta \!\left( \! x_p\!-\!\frac{q_1^++q_2^++q_3^+}{p_p^+} \right)
\!(2\pi)^3\,  4C_F\, \alpha_s\alpha_e \, \frac{1}{ p^+} \frac{1}{\xi_1 p^+}\frac{1}{\xi_2\bar\xi_1p^+}\, 
\nonumber\\
&&
\times
\,
(1+\bar\xi_1^2) \, (1+\bar\xi_2^2) 
\frac{1}{K_T^4}\frac{1}{q_1^2} \, 
\int_{b\bar b}e^{iP_T\cdot(\bar b-b)}
\Bigg\{
\bigg[ \xi_2^2-\frac{(1-2\xi_2)}{N_c^2-1}\bigg]
\bigg\langle \frac{1}{N_c} \tr\Big(\partial^iS_F(b)\partial^iS_F^\dagger({\bar b})\Big)
\bigg\rangle_T 
\nonumber\\
&&
\hspace{2cm}
-\frac{N_c^2}{N_c^2-1}\bigg\langle \frac{1}{N_c}\tr\Big(\partial^iS_F(b)S_F^\dagger({\bar b})\partial^iS_F({\bar b})S_F^\dagger(b)\Big)
s(b,\bar b)\bigg\rangle_T \Bigg\} \, \,  {\cal S}^{\rm Sud}(b,\bar{b}) \, ,
\label{finalwfwithsud}
\eeq
with
\beq
\label{Sudakov}
{\cal S}^{\rm Sud}(b,\bar{b})=\exp\left[ -\frac{\alpha_s}{2 \pi} \frac{C_A+C_F}{2} \ln^2\left(\frac{K_T^2(b-\bar{b})^2}{c_0^2}\right)  \right]
\eeq
and $c_0=2e^{-\gamma_E}$. With $|b-\bar{b}|\lesssim 1/Q_s$ determined by the averaged dipole scattering matrix $\Big\langle s(b,\bar{b})\Big\rangle_T$,
 and $|K_T|\gg Q_s$ in our kinematics, the modification due to the Sudakov factor is not evidently small. 
 
 Before we conclude this section, we would like to point out that our result in the back-to-back correlation limit can be written in terms of the transverse-momentum-dependent (TMD) gluon distributions. The first two TMD gluon distributions are defined as \cite{Marquet:2016cgx}
\beq
{\cal F}^{(1)}_{qg}(x_2,k_t)&=&\frac{4}{g^2}\int_{xy}e^{ik_t\cdot(x-y)}\left\langle\tr\left[\partial^iS_F(x)\partial^iS_F^\dagger(y)\right]\right\rangle_{x_2}, \\
{\cal F}^{(2)}_{qg}(x_2,k_t)&=&-\frac{4}{g^2}\int_{xy}e^{ik_t\cdot(x-y)}
\frac{1}{N_c}\Bigg\langle\tr\left[\partial^iS_F(x)S_F^\dagger(y)\partial^iS_F(y)S_F^\dagger(x)\right]\nonumber\\
&&
\hspace{5cm}
\times\, 
\tr\left[S_F(x)S_F^\dagger(y)\right]\Bigg\rangle_{x_2},
\eeq
where $\langle \cdots \rangle_{x_2}$ denotes average over the target boosted to rapidity $\ln {1/x_2}$.
Using these definitions of the TMD gluon distributions, we can write the production cross section at partonic level given in Eq. \eqref{finalwf} as 
\beq
\label{finalwf_TMDS}
&&
\hspace{-0.3cm}
(2\pi)^9\frac{d\sigma^{qA\to q\gamma g+X}}{d^3\q_1d^3\q_2d^3\q_3}= 
(2\pi)\delta(p^+-q_1^+-q_2^+-q_3^+) \, \alpha_s^2\alpha_e \, (2\pi)^3 \frac{1}{\xi_1p^+}\frac{1}{\xi_2\bar\xi_1p^+}
(1+\bar\xi_1^2)
(1+\bar\xi_2^2)
\, \nonumber\\
&&
\hspace{4cm}
\times \, 
\frac{1}{K_T^4}\frac{1}{q_1^2}\bigg\{
\left[\xi_2^2-\frac{\bar\xi_2^2}{N_c^2}\right]{\cal F}^{(1)}_{qg}(x_2,P_T)+{\cal F}^{(2)}_{qg}(x_2,P_T)\bigg\},
\eeq
which should be convoluted with quark distribution functions in order to arrive to the full production cross section as in Eq. \eqref{finalwfwithpdf}.

\section{The dilute target limit}
\label{Section:Dilute target}

The CGC cross section derived in Section~\ref{Section:WaveFunctional} allows us to study the dilute target limit which probes the linear (non-saturation) small-$x$ regime of the target. The interaction with the target in this limit is dominated by a single hard scattering of one of the propagating partons with the target
  and corresponds to the situation when the two hard jets are produced far from back-to-back, i.e., $|q_2|, |q_3|, |{q}_{2}+{q}_{3}|\sim |q_2-q_3|$.
The additional requirement of a collinear photon to be produced as well constrains the scattering off the dilute target to happen after the photon has been emitted. If the hard interaction with the target happens with the initial quark (before the emission of the photon) the probability of having a collinear photon in the final state is negligibly small. Neglecting the initial interaction amounts to setting $S_F(\omega)$ and $S_F^\dagger({\omega'})$ to a unit matrix in Eq.~(\ref{Partonic_X_section_1}). The cross section in this kinematics is then
\beq
&&
(2\pi)^9\frac{d\sigma^{qA\to q\gamma g+X}}{d^3\q_1d^3\q_2d^3\q_3} =  \frac{1}{N_c}g_s^2g_e^2 \;
(2\pi)\delta(p^+-q_1^+-q_2^+-q_3^+)  \, 
\frac{1}{\xi_1p^+} \, 
\frac{1}{\bar\xi_1\xi_2p^+}
\\
&&
\times
 \int_{y_1z_1,y_2z_2,y_3z_3,\omega v,\omega'  v'}
 \!\!\!\!\!\!\!\!\!\!\!\!\!\!\!\!\!\!\!\!\!\!\!\!\!\!\!\!
e^{iq_1\cdot(y_1-z_1)+iq_2\cdot(y_2-z_2)+iq_3\cdot(y_3-z_3)} \, 
\delta^{(2)}[\omega-(\bar\xi_1v+\xi_1z_1)] \, 
\delta^{(2)}[\omega'-(\bar\xi_1v'+\xi_1y_1)]\,
\nonumber\\
&&
\times \, 
\delta^{(2)}[v-(\bar\xi_2z_3+\xi_2z_2)]\, 
\delta^{(2)}[v'-(\bar\xi_2y_3+\xi_2y_2)]\, 
A^{\eta}(z_3-z_2) \,
A^{\eta}(y_3-y_2) \, 
(1+\bar\xi_1^2) \, (1+\bar\xi_2^2)
\nonumber\\
&&
\times \, 
\bigg\{{\cal A}_{\xi_2}^{\lambda}(v'-y_1) {\cal A}_{\xi_2}^{\lambda}(v-z_1) \, 
\bigg[ 
\frac{N_c^2}{2}
\bigg( s(z_2,y_2)Q(y_2,z_2,z_3,y_3)+ 1 -u(v)-u^\dagger(v')\bigg)\nonumber\\
&&
\hspace{6.5cm}
+
\frac{1}{2} \bigg( u(z_3)+u^\dagger(y_3)-s(z_3,y_3)-1\bigg)\bigg]\nonumber\\
&&
\hspace{0.5cm}
+ \, 
{A}^\lambda(v'-y_1){A}^\lambda(v-z_1)\bigg[
\frac{N_c^2-1}{2}\bigg( s(v,v')+1-u(v)-u^\dagger(v')  \bigg)\bigg]\nonumber\\
&&
\hspace{0.5cm}
- \, 
{\cal A}_{\xi_2}^\lambda(v'-y_1){A}^\lambda(v-z_1)\bigg[ 
\frac{N_c^2}{2}
\bigg( s(y_2,y_3)\Big[s(v,y_2)-u^\dagger(y_2)\Big] -u(v)+1\bigg)\nonumber\\
&&
\hspace{6.5cm}
+\frac{1}{2}\bigg( u^\dagger(y_3)+u(v)-s(v,y_3)-1\bigg)\bigg]\nonumber\\
&&
\hspace{0.5cm}
- \, 
{A}^\lambda(v'-y_1){\cal A}_{\xi_2}^\lambda(v-z_1)\bigg[ \frac{N_c^2}{2}\bigg( 
s(z_3,z_2)\Big[s(z_2,v')-u(z_2)\Big] -u^\dagger(v')+1\bigg)\nonumber\\
&&
\hspace{6.5cm}
+\, 
\frac{1}{2}\bigg( u(z_3)+u^\dagger(v')-s(z_3,v')-1\bigg)\bigg]\bigg\},\nonumber
\eeq
where we have defined $u(v)\equiv (1/N_c) \, \tr\, S_F(v)$. The terms involving a trace of a single Wilson line are somewhat unusual, but as we will see below they do not contribute in the kinematics we are interested in.
The integration over the photon transverse positions, and $\omega$, $v$, ${\omega'}$ and ${v'}$, gives
\beq
&&
(2\pi)^9\frac{d\sigma^{qA\to q\gamma g+X}}{d^3\q_1d^3\q_2d^3\q_3} =  \frac{1}{N_c}g_s^2g_e^2 \;
(2\pi)\delta(p^+-q_1^+-q_2^+-q_3^+)  \, 
\frac{(1+\bar\xi_1^2)}{\xi_1p^+} \, 
\frac{(1+\bar\xi_2^2)}{\bar\xi_1\xi_2p^+}
\\
&&
\times
 \int_{y_2z_2,y_3z_3}
e^{iq_1\cdot(v'-v)+iq_2\cdot(y_2-z_2)+iq_3\cdot(y_3-z_3)} \, 
{A}^{\eta}(z_3-z_2) \,
{A}^{\eta}(y_3-y_2) \, 
\nonumber\\
&&
\times \, 
\bigg\{\frac{ \xi_2 \, \bar{\xi_2}}{\xi_1}  \, |y_3 -y_2| \,|z_3 -z_2|\, K_1\left(\sqrt{\frac{\xi_2 \bar{\xi_2}}{\xi_1}} |y_3-y_2||q_1|\right)\,
K_1\left(\sqrt{\frac{\xi_2 \bar{\xi_2}}{\xi_1}} |z_3-z_2||q_1|\right)
\nonumber\\
&&
\hspace{0.5cm}  \times \bigg[ 
\frac{N_c^2}{2}
\bigg( s(z_2,y_2)Q(y_2,z_2,z_3,y_3)+ 1 -u(v)-u^\dagger(v')\bigg)+
\frac{1}{2} \bigg( u(z_3)+u^\dagger(y_3)-s(z_3,y_3)-1\bigg)\bigg]\nonumber\\
&&
+ \, 
 \frac{1}{q_1^2}\, \bigg[
\frac{N_c^2-1}{2}\bigg( s(v,v')+1-u(v)-u^\dagger(v')  \bigg)\bigg]\nonumber\\
&&
- \, 
 \sqrt{\frac{ \xi_2\, \bar{\xi_2}}{\xi_1}} \, \frac{1}{|q_1|}\,|y_3 -y_2| \, K_1\left(\sqrt{\frac{\xi_2 \bar{\xi_2}}{\xi_1}} |y_3-y_2||q_1|\right) \nonumber\\
&&
\hspace{0.5cm}  \times 
\bigg[ 
\frac{N_c^2}{2}
\bigg( s(y_2,y_3)\Big[s(v,y_2)-u^\dagger(y_2)\Big] -u(v)+1\bigg)+\frac{1}{2}\bigg( u^\dagger(y_3)+u(v)-s(v,y_3)-1\bigg)\bigg]\nonumber\\
&&
- \, 
 \sqrt{\frac{ \xi_2\, \bar{\xi_2}}{\xi_1 }} \, \frac{1}{|q_1|} \, |z_3 -z_2| \, K_1\left(\sqrt{\frac{\xi_2 \bar{\xi_2}}{\xi_1}} |z_3-z_2||q_1|\right)\nonumber\\
&&
\hspace{0.5cm}  \times \bigg[ \frac{N_c^2}{2}\bigg( 
s(z_3,z_2)\Big[s(z_2,v')-u(z_2)\Big] -u^\dagger(v')+1\bigg)
+\, 
\frac{1}{2}\bigg( u(z_3)+u^\dagger(v')-s(z_3,v')-1\bigg)\bigg]\bigg\},\nonumber
\eeq
where $v=\bar\xi_2z_3+\xi_2z_2$ and ${v'}=\bar\xi_2y_3+\xi_2y_2$. It is useful to change the integration coordinates to $v$ and ${v'}$, and $r=z_3 -z_2$ and ${r'}=y_3-y_2$. Then we have
\beq
&&
(2\pi)^9\frac{d\sigma^{qA\to q\gamma g+X}}{d^3\q_1d^3\q_2d^3\q_3} =  \frac{1}{N_c}g_s^2g_e^2 \;
(2\pi)\delta(p^+-q_1^+-q_2^+-q_3^+)  \, 
\frac{(1+\bar\xi_1^2)}{\xi_1p^+} \, 
\frac{(1+\bar\xi_2^2)}{\bar\xi_1 \xi_2p^+}
\\
&&
\times
 \int_{r r' v v'}
e^{iP_T\cdot({v'}-v)} \, e^{il_T\cdot({r'}-r)} \, 
{A}^{\eta}(r) \,
{A}^{\eta}(r') \, 
\nonumber\\
&&
\times \, 
\bigg\{\frac{ \xi_2 \, \bar{\xi_2}}{\xi_1}  \, |r| \,|r'|\, K_1\left(\sqrt{\frac{\xi_2 \bar{\xi_2}}{\xi_1}} |r||q_1|\right)\,
K_1\left(\sqrt{\frac{\xi_2 \bar{\xi_2}}{\xi_1}} |r'||q_1|\right)
\nonumber\\
&&
\hspace{0.5cm}  \times \bigg[ 
\frac{N_c^2}{2}
\bigg( s(z_2,y_2)Q(y_2,z_2,z_3,y_3)+ 1 -u(v)-u^\dagger(v')\bigg)+
\frac{1}{2} \bigg( u(z_3)+u^\dagger(y_3)-s(z_3,y_3)-1\bigg)\bigg]\nonumber\\
&&
+ \, 
 \frac{1}{q_1^2}\, \bigg[
\frac{N_c^2-1}{2}\bigg( s(v,v')+1-u(v)-u^\dagger(v')  \bigg)\bigg]\nonumber\\
&&
- \, 
 \sqrt{\frac{ \xi_2\, \bar{\xi_2}}{\xi_1}} \, \frac{1}{|q_1|} \, |r'| \, K_1\left(\sqrt{\frac{\xi_2 \bar{\xi_2}}{\xi_1}} |r'||q_1|\right) \nonumber\\
&&
\hspace{0.5cm}  \times 
\bigg[ 
\frac{N_c^2}{2}
\bigg( s(y_2,y_3)\Big[s(v,y_2)-u^\dagger(y_2)\Big] -u(v)+1\bigg)+\frac{1}{2}\bigg( u^\dagger(y_3)+u(v)-s(v,y_3)-1\bigg)\bigg]\nonumber\\
&&
- \, 
\sqrt{\frac{ \xi_2\, \bar{\xi_2}}{\xi_1}} \, \frac{1}{|q_1|} \, |r| \, K_1\left(\sqrt{\frac{\xi_2 \bar{\xi_2}}{\xi_1}} |r||q_1|\right)\nonumber\\
&&
\hspace{0.5cm}  \times \bigg[ \frac{N_c^2}{2}\bigg( 
s(z_3,z_2)\Big[s(z_2,v')-u(z_2)\Big] -u^\dagger(v')+1\bigg)
+\, 
\frac{1}{2}\bigg( u(z_3)+u^\dagger(v')-s(z_3,v')-1\bigg)\bigg]\bigg\},\nonumber
\eeq
where $P_T =q_1 + q_2 + q_3$ and $l_T = \xi_2 q_3 - \bar{\xi_2}q_2$. We have kept the old notations $z_2=v-\bar{\xi_2} r$, $z_3=v+ \xi_2 r$, $y_2={v'}- \bar{\xi_2}r'$ and $y_3={v'}+\xi_2 {r'}$ in the average for simplicity of the subsequent equations. For two hard jets with large momentum imbalance, both $|P_T|$ and $|l_T|$ are large. The terms in the curly brackets that do not depend on $v$, $v'$ or both, give $\delta^{(2)}(P_T)$ after integrating over $v$ and ${v'}$. This contribution is peaked around $|P_T|=0$ and can be dropped in the high-$|P_T|$ limit. The cross section simplifies to
\beq
&&
(2\pi)^9\frac{d\sigma^{qA\to q\gamma g+X}}{d^3\q_1d^3\q_2d^3\q_3} =  \frac{1}{N_c}g_s^2g_e^2 \;
(2\pi)\delta(p^+-q_1^+-q_2^+-q_3^+)  \, 
\frac{(1+\bar\xi_1^2)}{\xi_1p^+} \, 
\frac{(1+\bar\xi_2^2)}{\bar\xi_1\xi_2p^+}
\\
&&
\times
 \int_{r r' v v'}
e^{iP_T\cdot({v'}-v)} \, e^{il_T\cdot({r'}-r)} \, 
{A}^{\eta}(r) \,
{A}^{\eta}(r') \, 
\nonumber\\
&&
\times \, 
\bigg\{\frac{ \xi_2 \, \bar{\xi_2}}{\xi_1}  \, |r| \,|r'|\, K_1\left(\sqrt{\frac{\xi_2 \bar{\xi_2}}{\xi_1}} |r||q_1|\right)\,
K_1\left(\sqrt{\frac{\xi_2 \bar{\xi_2}}{\xi_1}} |r'||q_1|\right)
\nonumber\\
&&
\hspace{0.5cm}  \times \bigg[ 
\frac{N_c^2}{2}
 s(z_2,y_2)Q(y_2,z_2,z_3,y_3)-
\frac{1}{2} s(z_3,y_3)\bigg]
+ \, 
 \frac{1}{q_1^2}\, 
\frac{N_c^2-1}{2} \, s(v,v')\nonumber\\
&&
- \, 
\sqrt{\frac{ \xi_2\, \bar{\xi_2}}{\xi_1}} \, \frac{1}{|q_1|} \, |r'| \, K_1\left(\sqrt{\frac{\xi_2 \bar{\xi_2}}{\xi_1}} |r'||q_1|\right) 
\bigg[ 
\frac{N_c^2}{2}
\, s(y_2,y_3)\, s(v,y_2) -\frac{1}{2}\,s(v,y_3)\bigg]\nonumber\\
&&
- \, 
\sqrt{\frac{ \xi_2\, \bar{\xi_2}}{\xi_1 }} \, \frac{1}{|q_1|} \, |r| \, K_1\left(\sqrt{\frac{\xi_2 \bar{\xi_2}}{\xi_1}} |r||q_1|\right) \bigg[ \frac{N_c^2}{2}\,
s(z_3,z_2)\, s(z_2,v')
-\, 
\frac{1}{2}\, s(z_3,v')\bigg]\bigg\}.\nonumber
\eeq

In the dilute target limit, for both $|P_T|$ and $|l_T|$ large, one can expand the correlators for small transverse separations, which corresponds to expanding the Wilson lines to second order in the background field $A^-_a (z^+,{{z}})$ (two-gluon exchange). After expanding the modified Bessel functions of the second kind for small $|r|$ and $|r'|$, we obtain
\beq
&&
(2\pi)^9\frac{d\sigma^{qA\to q\gamma g+X}}{d^3\q_1d^3\q_2d^3\q_3} =  C_F g_s^2g_e^2 \;
(2\pi)\delta(p^+-q_1^+-q_2^+-q_3^+)  \, 
\frac{(1+\bar\xi_1^2)}{\xi_1p^+} \, 
\frac{(1+\bar\xi_2^2)}{\bar\xi_1\xi_2p^+}
\\
&&
\times
 \frac{1}{q_1^2}\,  \int_{r r' v v'}
e^{iP_T\cdot({v'}-v)} \, e^{il_T\cdot({r'}-r)} \, 
{A}^{\eta}(r) \,
{A}^{\eta}(r') \, 
\nonumber\\
&&
\times \, 
\bigg[
\,  
- g_s^2 N_c \Gamma  ({{z_2}}-{{y_2}}) - g_s^2 C_F \Gamma ({{z_3}}-{{y_3}})- g_s^2 C_F \Gamma ({{v}}-{{v'}}) \nonumber \\
&& \hspace{20pt} - \frac{g_s^2 N_c}{2} \left[ \Gamma ({{z_2}}-{{z_3}}) + \Gamma ({{y_2}}-{{y_3}}) - \Gamma ({{y_2}}-{{z_3}}) - \Gamma ({{z_2}}-{{y_3}}) \right]\nonumber\\
&&
\hspace{20pt} + \frac{g_s^2 N_c}{2} \Gamma ({{v}}-{{y_2}}) + \frac{g_s^2 N_c}{2} \Gamma ({{y_2}}-{{y_3}}) - \frac{g_s^2}{2 N_c} \Gamma ({{v}}-{{y_3}})\nonumber\\
&&
\hspace{20pt}  + \frac{g_s^2 N_c}{2} \Gamma ({{z_3}}-{{z_2}}) + \frac{g_s^2 N_c}{2} \Gamma ({{z_2}}-{{v'}}) - \frac{g_s^2}{2 N_c} \Gamma ({{z_3}}-{{v'}})\bigg],\nonumber
\label{eq:Amplitude_Gamma}
\eeq
where 
\be  
\Gamma  ({{z}} - {{y}}) = \int dx^+ \left[ {\tilde{\gamma}} (x^+, {{0}}) - {\tilde{\gamma}} (x^+, {{z-y}})\right]~,
\label{eq:Gammadef}
\ee
with ${\tilde{\gamma}} (x^+, {z-y})$ associated with the average value of the two-field correlator in the background field of the target:
\be 
\left \langle A^-_a (z^+,{{z}}) A^-_b(y^+,{{y}}) \right\rangle_T= \delta ^{ab} \delta (z^+ - y^+) {\tilde{\gamma}} (z^+, {{z}} - {{y}})~.
\ee
After performing the Fourier transformations in Eq. (\ref{eq:Amplitude_Gamma}), we get
\beq
&&
(2\pi)^9\frac{d\sigma^{qA\to q\gamma g+X}}{d^3\q_1d^3\q_2d^3\q_3} =  C_F g_s^4g_e^2 \;
(2\pi)\delta(p^+-q_1^+-q_2^+-q_3^+)  \, 
\frac{(1+\bar\xi_1^2)}{2\xi_1p^+} \, 
\frac{(1+\bar\xi_2^2)}{2\bar\xi_1\xi_2p^+}\, \frac{1}{q_1^2}\, \, S_\perp \frac{f(P_T)}{P_T^2} \,\,\nonumber
\\
&&
\times \,
\bigg[
\, 
 N_c \, \frac{1}{m_T^2}
  + C_F \, \frac{1}{n_T^2}
 + \, 
C_F \, \frac{1}{l_T^2}  + N_c \, \frac{n_T \cdot m_T}{n_T^2 m_T^2} 
 - \, 
N_c \, \frac{l_T \cdot m_T}{l_T^2 m_T^2}
 - \frac{1}{ N_c}\frac{l_T \cdot n_T}{l_T^2 n_T^2} 
 \bigg].
\eeq
The newly introduced transverse momenta in the above expression are defined as $m_T = \bar{\xi_2}q_1 + q_3$ and $n_T = \xi_2 q_1 +q_2$. The factor of transverse area of the target, $S_\perp$ arises as the result of the impact parameter integration. We have also introduced $f(P_T)$ as
\be 
f(P_T) \equiv - P_T^2 \int d^2 {{r}} \, \Gamma ({{r}}) e^{-i P_T \cdot {{r}}} = P_T^2 \int dx^+ {\tilde{\gamma}}  (x^+, P_T)~,
\ee
which can be related to the dipole scattering amplitude, i.e., to the unintegrated gluon distribution appearing in the total cross section for deep inelastic scattering~\cite{Iancu:2013dta,Kotko:2015ura}:
\be
f(P_T) =\frac{1}{(2\pi) \alpha_s S_\perp}\frac{N_c}{N_c^2-1} P_T^2\int_{ {{b'}} {\bar{b'}}}\
e^{i P_T \cdot ({\bar{b'}}-{{b'}})}\left \langle s ({{b'}},{\bar{b'}})\right\rangle_T\ .
\ee
The final result for proton-nucleus scattering convoluted with the quark distribution in the proton can be written
\beq
\label{final_dilute_target}
&&
(2\pi)^9\frac{d\sigma^{pA\to q\gamma g+X}}{d^3\q_1d^3\q_2d^3\q_3}=
\int dx_p \,x_p\, f^q_{\mu^2}(x_p) \ \delta\left(  x_p-\frac{q_1^++q_2^++q_3^+}{p_p^+} \right)\nonumber\\
&&\times\ 
4(2\pi)^3\, \alpha_s\alpha_e \, \frac{1}{ p^+} \frac{(1+\bar\xi_1^2)}{\xi_1p^+} \, 
\frac{(1+\bar\xi_2^2)}{\bar\xi_1\xi_2p^+}\, \frac{1}{q_1^2}
\nonumber\\
&&
\times 
\bigg[
\, 
 N_c \, \frac{1}{m_T^2}
  + C_F \, \frac{1}{n_T^2}
 + \, 
C_F \, \frac{1}{l_T^2}  + N_c \, \frac{n_T \cdot m_T}{n_T^2 m_T^2} 
 - \, 
N_c \, \frac{l_T \cdot m_T}{l_T^2 m_T^2}
 - \frac{1}{ N_c}\frac{l_T \cdot n_T}{l_T^2 n_T^2} 
 \bigg]\, 
 \nonumber\\
&&
\times
\int_{b'\bar b'}e^{iP_T\cdot(\bar b'-b')}\left\langle s(b',\bar b')\right\rangle_T\ .
\label{Eq:finaldilute}
\eeq

There is no ordering between the transverse momentum of the jets and their momentum imbalance. The contribution from the Sudakov logarithms is therefore small.

\section{Discussion}
\label{Section:Discussion}

In conclusion, we have computed the cross section for production of a soft photon and two hard jets in the hybrid formalism suitable for the forward rapidity region in p-A collisions. After obtaining the full cross section, Eq.~\eqref{Partonic_X_section_final}, we have calculated two limits of this production cross section: the back-to-back correlation limit (the final result is given in Eq.~\eqref{finalwfwithpdf}) and the dilute target limit (the final result is given in Eq.~\eqref{final_dilute_target})

In the back-to-back correlation limit, the produced jets have transverse momenta much larger than the saturation scale of the target whereas the transverse momentum imbalance of the jets is of the order of the saturation scale. We have shown that the full production cross section in this limit simplifies and it can be written in terms of the transverse-momentum-dependent (TMD) gluon distributions ${\cal F}^{(1)}_{qg}(x_2,k_t)$ and ${\cal F}^{(2)}_{qg}(x_2,k_t)$. Our expression in the back-to-back correlation limit is very similar to the one for forward djiet production in the same limit \cite{Marquet:2016cgx} (up to the kinematical factors due to the emission of the extra soft photon), which coincides with the small-x limit of the TMD formula \cite{Dominguez:2011wm,Marquet:2016cgx} in their overlapping validity region. Obviously, the production cross section of the soft photon and two hard jets is suppressed by a power of $\alpha_{em}$ compared to the forward dijet production cross section, but it is enhanced by the inverse of the transverse momentum of the soft photon in the back-to-back correlation limit. Thus, the $\alpha_{em}$ suppression can be compensated by the transverse momenta of the soft photon which indicates that this observable might be also very interesting experimentally.  Moreover, forward dijet production was studied to show the agreement between the CGC and TMD frameworks \cite{Marquet:2016cgx}. Our results have shown that the emission of the soft photon does not spoil the TMD structure that was seen in the forward dijet production. Finally, we would like to point out that our result shows sensitivity to saturation scale and its evolution in rapidity is given by the  nonlinear Jalilian-Marian--Iancu--McLerran--Weigert--Leonidov--Kovner \cite{JIMWLK} or Balitsky-Kovchegov \cite{balitsky,Kovchegov} equations. 

\begin{figure}[hbt]
\begin{center}
\includegraphics[width=\textwidth]{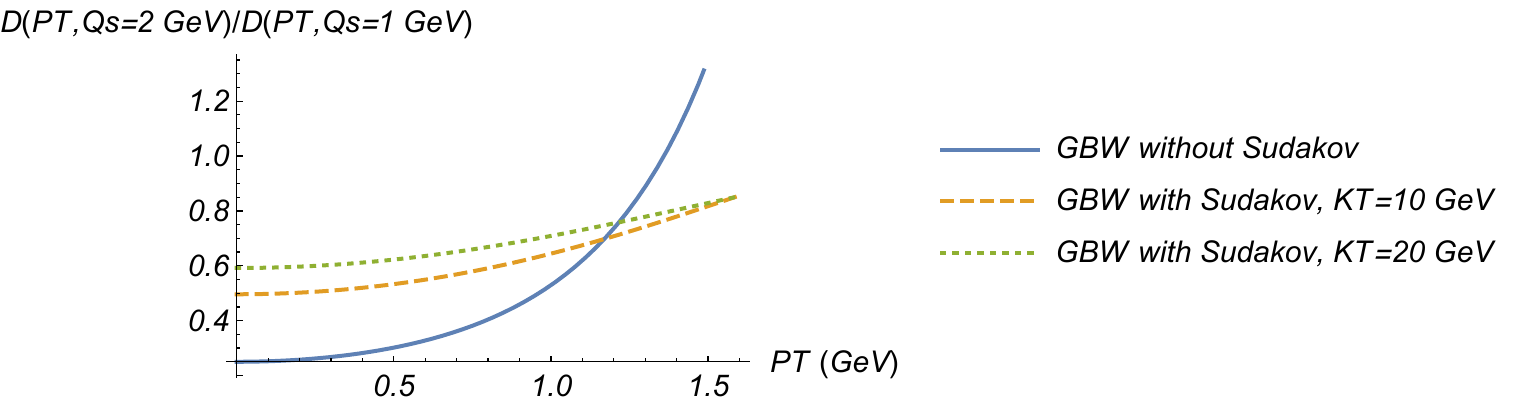}
\end{center}
\caption{Ratios of Eq. \eqref{dipolegbw} (solid blue line), Eq. \eqref{dipoleS} with $|K_T|=10$ GeV (dashed orange line) and Eq. \eqref{dipoleS} with $|K_T|=20$ GeV (dotted green line) with $Q_s=2$ GeV over the same equations with  $Q_s=1$ GeV, versus $|P_T|$.}
\label{fig:Sud}
\end{figure}

\begin{figure}[hbt]
\begin{center}
\includegraphics[width=\textwidth]{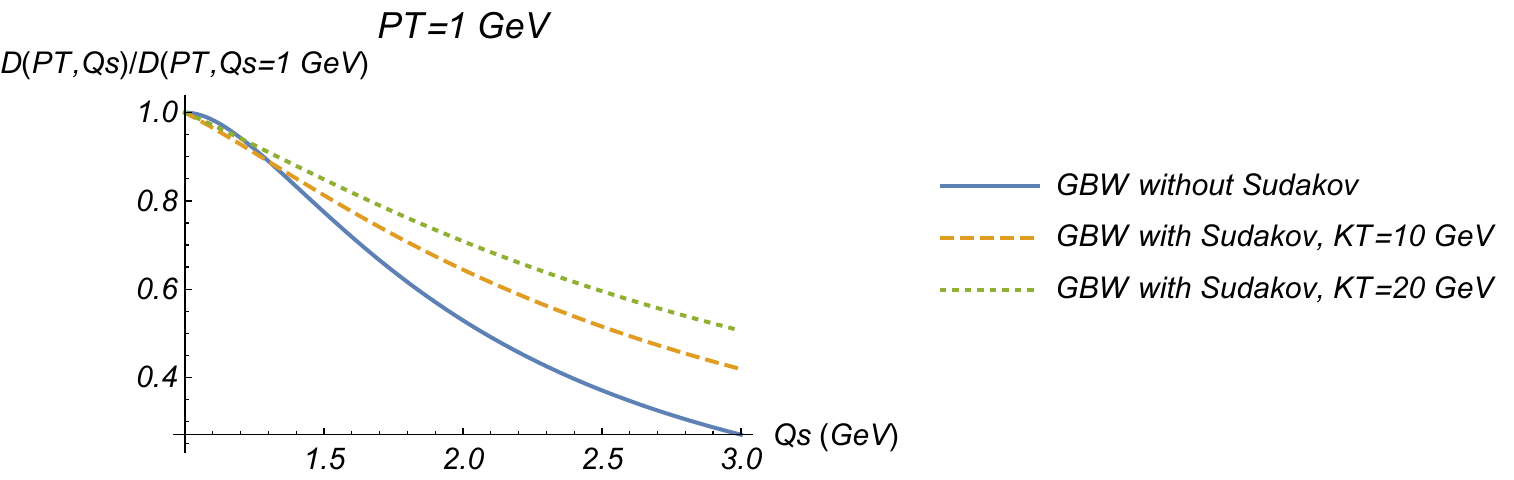}
\end{center}
\caption{Eq. \eqref{dipolegbw} (solid blue line), Eq. \eqref{dipoleS} with $|K_T|=10$ GeV (dashed orange line) and Eq. \eqref{dipoleS} with $|K_T|=20$ GeV (dotted green line) normalised to their values with  $Q_s=1$ GeV, versus $Q_s$, for $|P_T|=1$ GeV.}
\label{fig:Sud2}
\end{figure}

Unfortunately (unlike for some other observables, see e.g. \cite{Altinoluk:2015dpi}), the sensitivity to the saturation scale is somewhat washed out by the corrections due to the Sudakov emission.
In order to roughly estimate such effect for the piece containing ${\cal F}^{(1)}_{qg}(x_2,k_t)$,
let us take  the Golec-Biernat--W\"usthoff (GBW) model \cite{gbw} for $\left\langle s(b',\bar b')\right\rangle_T$ which corresponds to
\beq
{\cal D}(P_T)\propto \frac{P_T^2\  e^{-P_T^2/Q_s^2}}{Q_s^2}\ ,
\label{dipolegbw}
\eeq
with $Q_s$ the saturation scale of the target. The model is not realistic for large $P_T$ but it suits our purpose, since as explained above our main interest is in the small momentum transferred from the target. In Fig. \ref{fig:Sud} we show the ratio of Eqs. \eqref{dipolegbw} for two different values of $Q_s=1$ and 2 GeV versus $|P_T|$. In Fig. \ref{fig:Sud2} we show Eq. \eqref{dipolegbw} versus $Q_s$ for $|P_T|=1$ GeV, normalised to its value at $Q_s=1$ GeV.
Now, to estimate the effect of the Sudakov, we define
\beq
{\cal D}_{Sud}(P_T)=\frac{N_c P_T^2}{2 \pi^2 \alpha_s} \frac{1}{(2\pi)^2}\int_{b'\bar b'}e^{iP_T\cdot(\bar b'-b')} \, \left \langle s(b',\bar b')\right \rangle_T\, {\cal S}^{\rm Sud}(b',\bar{b}')
\label{dipoleS}
\eeq
and plot in Fig. \ref{fig:Sud} the ratio of  Eqs. \eqref{dipoleS}   for $|K_T|=10$ and 20 GeV for two different values of $Q_s=1$ and 2 GeV versus $|P_T|$, while in Fig. \ref{fig:Sud2} we show Eq. \eqref{dipoleS} versus $Q_s$ for $|P_T|=1$ GeV, normalised to its value at $Q_s=1$ GeV. We have used $\alpha_s=0.2$, and the SU(3) values for  $C_A$ and $C_F$. 
While the Sudakov factor somewhat washes away the effect of saturation,  the sensitivity remains\footnote{The region of large $|P_T|> Q_s$ cannot be trusted in the GBW model that produces a Gaussian decrease in contrast with the expected perturbative power-law behavior.}  in the region $|P_T|\simeq Q_s$ even up to fairly high values of $|K_T|$. This conclusion may be of more general applicability for other processes studied in \cite{Mueller:2013wwa}.

We have also discussed the dilute target limit of the production cross section of a soft photon and two hard jets. In this limit, the momentum imbalance between the two produced jets is large and our result simplifies to a single dipole gluon distribution of the target which follows the linear Balitsky-Fadin-Kuraev-Lipatov evolution \cite{bfkl}, as opposed to the back-to-back correlation limit.

The computation in this work constitutes a first step towards photon-jet production at NLO and eventually a complete NLO calculation of photon production in the hybrid formalism.

\section*{Acknowledgments}
TA expresses his gratitude to the Heavy Ion Phenomenology group at LIP, Lisboa, and the Departamento de F\'{\i}sica de Part\'{\i}culas at Universidade de Santiago de Compostela, for  support when part of this work was done.
NA was supported by the European Research Council grant HotLHC ERC-2011-StG-279579, Ministerio de Ciencia e Innovaci\'on of Spain under project FPA2014-58293-C2-1-P and Unidad de Excelencia Mar\'{\i}a de Maetzu under project MDM-2016-0692,  Xunta de Galicia (Conseller\'{\i}a de Educaci\'on) within the Strategic Unit AGRUP2015/11, and FEDER. 
AK was supported by the NSF Nuclear Theory grant 1614640, the Fulbright US scholar program and the CERN scientific associateship. ML was supported
by the  Israeli Science Foundation grants \# 1635/16 and \# 147/12;  AK and ML  were also supported by the
BSF grants \#2012124 and \#2014707. 
EP is supported by the European Community under the ``Ideas'' programme QWORK (contract no. 320389).
The work of TA, NA, ML and EP has been performed in the framework of   the COST Action CA15213 THOR.

 \appendix

\section{Details of the calculation of dressed quark state}
\label{Section:AppendixA}

\subsection{Dressed quark state in the mixed space}
\label{Section:Dressed_Quark_in_Mixed_State}

Even though we are considering incoming quark with vanishing transverse momenta in this paper, for future work we will calculate the mixed space expression for the dressed quark state with non vanishing transverse momentum of the incoming quark. For this general case, the full momentum expression of the dressed quark in terms of the bare states can be written as 
\beq
&&\left| (\rmq) [p^+,p]^\alpha_s\right\rangle_D= 
A^q\left| (\rmq) [p^+,p]_s^\alpha\right\rangle_0\nonumber \\ 
&+&
A^{q\gamma} \; g_e\sum_{s',\lambda} \int \frac{dk_1^+}{2\pi} \frac{d^2k_1}{(2\pi)^2} \; 
F_{(\rmq\rmp)}^{(1)}\left[ (\rmp)[k_1^+,k_1]^\lambda, (\rmq)[p^+-k_1^+,p-k_1]_{ss'}\right]\nonumber\\
&&
\hspace{6cm}
\times
\left| (\rmq) \left[p^+-k_1^+,p-k_1\right]_{s'}^\alpha ; (\rmp) [k_1^+,k_1]_\lambda \right\rangle_0 \nonumber\\
&+&A^{qg} \; g_s\sum_{s',\eta} \int \frac{dk_2^+}{2\pi}\frac{d^2k_2}{(2\pi)^2} \, t^c_{\alpha\beta} \; 
F^{(1)}_{(\rmq\rmg)}\left[ (\rmg)[k_2^+,k_2]^\eta, (\rmq)[p^+-k_2^+,p-k_2]_{ss'}\right] \nonumber\\
&&
\hspace{6cm}
\times
\left| (\rmq)\left[p^+-k_2^+,p-k_2 \right]_{s'}^\beta; (\rmg)\left[k_2^+,k_2\right]_{\eta}^c \right\rangle_0 \nonumber\\
&+&
A^{qg\gamma}\; g_sg_e\sum_{s's''}\sum_{\lambda\eta} \int \frac{dk_1^+}{2\pi}\frac{d^2k_1}{(2\pi)^2}\frac{dk_2^+}{2\pi}\frac{d^2k_2}{(2\pi)^2} \; t^c_{\alpha\beta}\nonumber\\
&&
\times
\Bigg\{ 
F^{(2)}_{(\rmq\rmp-\rmq\rmg)}\left[ (\rmp)[k_1^+,k_1]^\lambda , (\rmg)[k_2^+,k_2]^\eta , (\rmq)[p^+-k_1^+-k_2,p-k_1-k_2]_{ss''} \right]
\nonumber\\
&&
\hspace{0.2cm}
+ \; 
F^{(2)}_{(\rmq\rmg-\rmq\rmp)}\left[ (\rmg)[k_2^+,k_2]^\eta , (\rmp)[k_1^+,k_1]^\lambda , (\rmq)[p^+-k_2^+-k_1^+,p-k_2-k_1]_{ss''}\right]
\Bigg\}\nonumber\\
&&
\hspace{1.5cm}
\times
\left| (\rmq) \left[ p^+-k_1^+-k_2^+,p -k_1-k_2\right]_{s''}^\beta , (\rmg) \left[k_2^+,k_2 \right]^c_\eta , (\rmp) \left[k_1^+,k_1\right]^\lambda \right\rangle_0 \, .
\label{dwf1p}
\eeq
As it is argued in Section \ref{Section:WaveFunctional}, for our purposes we can set the normalizations $A^q, A^{q\gamma}, A^{qg}$ and $A^{qg\gamma}$ to one. The explicit momentum space expressions of the splitting amplitudes $F^{(1)}_{(\rmq\rmp)}$ and $F^{(2)}_{(\rmq\rmp-\rmq\rmg)}$ are given in Eqs. \eqref{F1} and \eqref{F2} respectively. Now, let us consider each component separately. 

\noindent (i) {\it bare quark component:} It is completely trivial to write the bare quark component of the dressed state in the mixed space:
\beq
\label{trivial}
\left|(\rmq)[p^+,p]_s^{\alpha}\right\rangle_0=\int_\omega e^{-ip\cdot\omega}\left|(\rmq)[p^+,\omega]_s^{\alpha}\right\rangle_0\,.
\eeq  

\noindent (ii) {\it bare quark-photon component:} The bare quark-photon component of the dressed quark can be written as 
\beq
&&
g_e\sum_{s'\lambda}\int\frac{dk_1^+}{2\pi}\frac{d^2k_1}{(2\pi)^2} F_{(\rmq\rmp)}^{(1)}\left[ (\rmp)[k_1^+,k_1]^\lambda, (\rmq)[p^+-k_1^+,p-k_1]_{ss'}\right] 
\nonumber\\
&&
\hspace{6cm}
\times
\left| (\rmq) \left[p^+-k_1^+,p-k_1\right]_{s'}^\alpha ; (\rmp) [k_1^+,k_1]_\lambda \right\rangle_0 \nonumber\\
&&=
g_e\sum_{s'\lambda}\int\frac{dk_1^+}{2\pi}\frac{d^2k_1}{(2\pi)^2}\int_{z_1z_2x_1v}e^{-ik_1\cdot(z_1+x_1)-i(p-k_1)\cdot(z_2+v)}
\nonumber\\
&&
\times
F^{(1)}_{(\rmq\rmp)}\left[(\rmp)[k_1^+,z_1]^{\lambda}; (\rmq)[p^+-k_1^+,z_2]_{ss'}\right] 
\left| (\rmq)[p^+-k_1^+,v]_{s'}^{\alpha}; (\rmp)[k_1^+,x_1]^\lambda\right\rangle_0\,.
\eeq
Integration over $k_1$ results in  $\delta^{(2)}[z_1-(z_2+v-x_1)]$. After trivially integrating over $z_1$ by using the $\delta$-function, and renaming $z_2+v=\omega$, we get
\beq
&&
g_e\sum_{s'\lambda}\int\frac{dk_1^+}{2\pi}\frac{d^2k_1}{(2\pi)^2} F_{(\rmq\rmp)}^{(1)}\left[ (\rmp)[k_1^+,k_1]^\lambda, (\rmq)[p^+-k_1^+,p-k_1]_{ss'}\right] 
\nonumber\\
&&
\hspace{6cm}
\times
\left| (\rmq) \left[p^+-k_1^+,p-k_1\right]_{s'}^\alpha ; (\rmp) [k_1^+,k_1]_\lambda \right\rangle_0 \nonumber\\
&&= g_e\sum_{s'\lambda}\int \frac{dk_1^+}{2\pi}\int_{\omega vx_1}e^{-ip\cdot\omega}
F^{(1)}_{(\rmq\rmp)}\left[(\rmp)[k_1^+,\omega-x_1]^\lambda; (\rmq)[p^+-k_1^+,\omega-v]_{ss'}\right]\nonumber\\
&&
\hspace{6cm}
\times
\left| (\rmq) [p^+-k_1^+,v]_{s'}^\alpha ; (\rmp) [k_1^+,x_1]^\lambda \right\rangle_0\,.
\eeq
Now, we need to calculate the Fourier transform of the splitting amplitude by using its explicit expression in full momentum space,
\beq
F^{(1)}_{(\rmq\rmp)}\left[(\rmp)[k_1^+,\omega-x_1]^\lambda; (\rmq)[p^+-k_1^+,\omega-v]_{ss'}\right]&=&
\int\frac{d^2q_1}{(2\pi)^2} \frac{d^2q}{(2\pi)^2} e^{iq_1\cdot(\omega-x_1)+i(q-q_1)\cdot(\omega-v)}\nonumber\\
&&
\times
\left[\frac{-\phi^{\lambda\bar\lambda}_{ss'}(\xi_1)}{\sqrt{2\xi_1p^+}}\right]
\frac{(\xi_1q-q_1)^{\bar\lambda}}{(\xi_1q-q_1)^2}\ .
\eeq
After performing the following change of variables:
\beq
\xi_1q-q_1&=&\xi_1P,\\
q_1&=&K,
\eeq
it is straightforward to perform the integrations over $P$ and $K$. The mixed space expression of the splitting amplitude reads
\beq
\label{F1_mixed}
&&
F^{(1)}_{(\rmq\rmp)}\left[(\rmp)[k_1^+,\omega-x_1]^\lambda; (\rmq)[p^+-k_1^+,\omega-v]_{ss'}\right]=\\
&&
\hspace{6cm}
=
\bigg[\frac{(-i)}{\sqrt{2\xi_1p^+}}\phi^{\lambda\bar\lambda}_{ss'}(\xi_1)\bigg]A^{\bar\lambda}(v-x_1)\delta^{(2)}[\omega\!-\!(\bar\xi_1v+\xi_1x_1)].\nonumber
\eeq
Finally, by using Eq. \eqref{F1_mixed}, we can write the bare quark-photon component of the dressed quark state in mixed space as 
\beq
\label{qgamma_component}
&&
g_e\sum_{s'\lambda}\int\frac{dk_1^+}{2\pi}\frac{d^2k_1}{(2\pi)^2} F_{(\rmq\rmp)}^{(1)}\left[ (\rmp)[k_1^+,k_1]^\lambda, (\rmq)[p^+-k_1^+,p-k_1]_{ss'}\right] 
\nonumber\\
&&
\hspace{6cm}
\times
\left| (\rmq) \left[p^+-k_1^+,p-k_1\right]_{s'}^\alpha ; (\rmp) [k_1^+,k_1]_\lambda \right\rangle_0 \nonumber\\
&&
=g_e\sum_{s'\lambda}\int\frac{dk_1^+}{2\pi}\int_{\omega v x_1}e^{-ip\cdot\omega} 
\bigg[\frac{(-i)}{\sqrt{2\xi_1p^+}}\phi^{\lambda\bar\lambda}_{ss'}(\xi_1)\bigg] 
A^{\bar\lambda}(v-x_1) )\delta^{(2)}[\omega\!-\!(\bar\xi_1v+\xi_1x_1)] \nonumber\\
&&
\hspace{6cm}
\times
\left| (\rmq) [p^+-k_1^+,v]_{s'}^\alpha ; (\rmp) [k_1^+,x_1]^\lambda \right\rangle_0\,.
\eeq

\noindent (iii) {\it bare quark-gluon component}: The calculation of the bare quark-gluon component is exactly the same as the bare quark-photon component. Thus we can read off the result from the final expression of the bare quark-photon component:
\beq
\label{qg_component}
&&
g_s\sum_{s'\eta}\int\frac{dk_2^+}{2\pi}\frac{d^2k_2}{(2\pi)^2} \, t^c_{\alpha\beta} \, F_{(\rmq\rmg)}^{(1)}\left[ (\rmg)[k_2^+,k_2]^\eta, (\rmq)[p^+-k_2^+,p-k_2]_{ss'}\right] 
\nonumber\\
&&
\hspace{6cm}
\times
\left| (\rmq) \left[p^+-k_2^+,p-k_2\right]_{s'}^\beta ; (\rmg) [k_2^+,k_2]^c_{\eta} \right\rangle_0 \nonumber\\
&&
=g_s\sum_{s'\eta}\int\frac{dk_2^+}{2\pi}\int_{\omega v x_2}e^{-ip\cdot\omega}  t^{c}_{\alpha\beta}
\bigg[\frac{(-i)}{\sqrt{2\xi_2p^+}} \phi^{\eta\bar\eta}_{ss'}(\xi_2)\bigg] 
\bar{A}^{\bar\eta}_{\xi_2}(v-x_2) \delta^{(2)}[\omega\!-\!(\bar\xi_2v+\xi_2x_2)] \nonumber\\
&&
\hspace{6cm}
\times
\left| (\rmq) [p^+-k_2^+,v]_{s'}^\beta ; (\rmg) [k_2^+,x_2]^c_{\eta}\right\rangle_0\,.
\eeq
The 
Weizs\''acker-Williams field in the quark-gluon splitting function takes into account the Ioffe time restriction. It reads
\beq
\label{Modified_A}
\bar{A}^{\bar\eta}_{\xi}(v-{x_1})=-\frac{1}{2\pi}\frac{(v-{x_1})^{\bar\eta}}{(v-{x_1})^2}\left[1-J_0\left(|v-{x_1}|\sqrt{2\xi(1-\xi)\frac{p^+}{\tau}}\right)\right].
\eeq
The Ioffe time restriction ensures that the life time of the quark-gluon pair is larger than the propagation time of the pair through the target. Hence, it guarantees that only quark-gluon pairs that are resolved by the target during the interaction are included in the projectile wave function. In the definition of the modified Weizs\''acker-Williams, Eq.\eqref{Modified_A}, $\tau$ can be identified as longitudinal size of the target at the initial energy.
\\

\noindent (iv) {\it bare quark-gluon-photon component} : The two terms in this component, as explained previously, correspond to two different orderings of the emissions. Let us first consider the first term where the photon is emitted before the gluon:
\beq
&&
\hspace{-1cm}
g_sg_e\sum_{s''\lambda\eta}\int \frac{dk_1^+}{2\pi} \frac{dk_2^+}{2\pi} \frac{d^2k_1}{(2\pi)^2} \frac{d^2k_2}{(2\pi)^2} t^c_{\alpha\beta} \nonumber\\
&&
\times
F^{(2)}_{(\rmq\rmp-\rmq\rmg)}\left[ (\rmp)[k_1^+,k_1]^\lambda; (\rmg)[k_2^+,k_2]^c_\eta, (\rmq)[p^+-k_1^+-k_2^+, p-k_1-k_2]_{ss''} \right]
\nonumber\\
&&
\times
\left| (\rmq)[p^+-k_1^+-k_2^+,p-k_1-k_2]^\beta_{ss''}; (\rmg)[k^+_2,k_2]^c_{\eta}; (\rmp)[k_1^+,k_1]^{\lambda}\right\rangle_0
\nonumber\\
&&
\hspace{-1cm}
=
g_sg_e\sum_{s''\lambda\eta}\int \frac{dk_1^+}{2\pi} \frac{dk_2^+}{2\pi} \frac{d^2k_1}{(2\pi)^2} \frac{d^2k_2}{(2\pi)^2} t^c_{\alpha\beta}
\int_{x_iz_j}e^{-ik_1\cdot(z_1+x_1)-ik_2\cdot(z_2+x_2)-i(p-k_1-k_2)\cdot(z_3+x_3)}\nonumber\\
&&
\times
F^{(2)}_{(\rmq\rmp-\rmq\rmg)}\left[ (\rmp)[k_1^+,z_1]^\lambda; (\rmg)[k_2^+,z_2]^c_\eta, (\rmq)[p^+-k_1^+-k_2^+, z_3]_{ss''} \right]
\nonumber\\
&&
\times
\left| (\rmq)[p^+-k_1^+-k_2^+,x_3]^\beta_{ss''}; (\rmg)[k^+_2,x_2]^c_{\eta}; (\rmp)[k_1^+,x_1]^{\lambda}\right\rangle_0\,.
\eeq
Here we have introduced a compact notation $\int_{x_iz_j}\equiv\int_{x_1x_2x_3z_1z_2z_3}$. Similar to the previous components, we can now integrate over $k_1$ and $k_2$. These two integrations give two $\delta$-functions that can be used to integrate over the variables $z_1$ and $z_2$. Finally, similar to the quark-photon and quark-gluon components, we can perform a change of variables for $z_3\to \omega-x_3$ to write down the mixed space expression of the first term in the bare quark-photon-gluon component of the dressed quark state  as 
\beq
&&
g_sg_e\sum_{s''\lambda\eta}\int \frac{dk_1^+}{2\pi} \frac{dk_2^+}{2\pi} t^{c}_{\alpha\beta}\int_{wx_1x_2x_3}e^{-ip\cdot\omega}\nonumber\\
&&
\times
F^{(2)}_{(\rmq\rmp-\rmq\rmg)}\left[ (\rmp)[k_1^+,\omega-x_1]^\lambda; (\rmg)[k_2^+,\omega-x_2]^c_\eta, (\rmq)[p^+-k_1^+-k_2^+, \omega-x_3]_{ss''} \right]\nonumber\\
&&
\times
\left| (\rmq)[p^+-k_1^+-k_2^+,x_3]^\beta_{ss''}; (\rmg)[k^+_2,x_2]^c_{\eta}; (\rmp)[k_1^+,x_1]^{\lambda}\right\rangle_0\,.
\eeq
Now, let us calculate the Fourier transform of the splitting amplitude $F^{(2)}_{(\rmq\rmp-\rmq\rmg)}$ by using its explicit expression in full momentum space, Eq. \eqref{F2}:
\beq
&&
F^{(2)}_{(\rmq\rmp-\rmq\rmg)}\left[ (\rmp)[k_1^+,\omega-x_1]^{\lambda}; (\rmg)[k_2^+,\omega-x_2]^\eta; (\rmq)[p^+-k_1^+-k_2^+,\omega-x_3]_{ss''}\right]\nonumber\\
&&=\sum_{s'}\
\frac{\phi^{\lambda\bar\lambda}_{ss'}(\xi_1)}{\sqrt{2\xi_1p^+}}
\frac{\tilde{\phi}^{\eta\bar\eta}_{s's''}(\xi_1,\xi_2)}{\sqrt{2\xi_2p^+}}
\int\frac{d^2p}{(2\pi)^2}\frac{d^2k_1}{(2\pi)^2}\frac{d^2k_2}{(2\pi)^2}e^{ik_1\cdot(w-x_1)+ik_2\cdot(w-x_2)+i(p-k_1-k_2)\cdot(\omega-x_3)}\nonumber\\
&&
\times
\frac{(\xi_1p-k_1)^{\bar\lambda}}{(\xi_1p-k_1)^2}
\frac{ [\xi_2(p-k_1)-\bar\xi_1k_2]^{\bar\eta} }{ \xi_2(\xi_1p-k_1)^2 + \xi_1(\xi_2p-k_2)^2 - (\xi_2k_1-\xi_1k_2)^2}\ .
\eeq
After performing the following change of variables:
\beq
\xi_1p-k_1&=&P,\\
\xi_2(p-k_1)-\bar\xi_1k_2&=&\bar\xi_1K,
\eeq 
we can integrate over $p$ to get the following expression:
\beq
&&
F^{(2)}_{(\rmq\rmp-\rmq\rmg)}\left[ (\rmp)[k_1^+,\omega-x_1]^{\lambda}; (\rmg)[k_2^+,\omega-x_2]^\eta; (\rmq)[p^+-k_1^+-k_2^+,\omega-x_3]_{ss''}\right]\nonumber\\
&&=\sum_{s'}
\frac{\phi^{\lambda\bar\lambda}_{ss'}(\xi_1)}{\sqrt{2\xi_1p^+}}
\frac{\tilde{\phi}^{\eta\bar\eta}_{s's''}(\xi_1,\xi_2)}{\sqrt{2\xi_2p^+}}
\delta^{(2)}\left[\omega-\left(\xi_1x_1+\xi_2x_2+(1-\xi_1-\xi_2)x_3\right)\right]\nonumber\\
&&
\times
\int \frac{d^2P}{(2\pi)^2}\frac{d^2K}{(2\pi)^2}e^{-iP\cdot\frac{(\omega-x_1)}{\bar \xi_1}-iK\cdot(x_3-x_2)}\frac{1}{\xi_1}\frac{P^{\bar\lambda}}{P^2}\frac{K^{\bar\eta}}{K^2+\frac{\xi_2(1-\xi_1-\xi_2)}{\xi_1{\bar\xi_1}^2}P^2}\ .
\eeq
Now, we can perform the integrals over $P$ and $K$ by considering the following integral 
\beq
\label{F2_FT_with_ioffe}
\int \frac{d^2P}{(2\pi)^2} \frac{d^2K}{(2\pi)^2} e^{iP\cdot r+iK\cdot r'}\frac{P^l}{P^2}\frac{K^m}{K^2+c_0P^2}\ .
\eeq

In order to be consistent with the ''{\it bare quark-gluon component} '' we should introduce the Ioffe time restriction for the gluon emission. Effectively, this corresponds to restricting the integration region of the momentum $K$, which is conjugate to the quark-gluon dipole size, from above, i.e.,
\beq
K^2 < 2\frac{\xi_2}{\bar \xi_1}\bigg( 1-\frac{\xi_2}{\bar \xi_1}\bigg)\frac{p^+-k_1^+}{\tau}\equiv \alpha.
\eeq
Here, $\tau$ corresponds to longitudinal size of the target at some given initial energy. We can perform the integral over $P$ in Eq.\eqref{F2_FT_with_ioffe} which simply reads
\beq
\int \frac{d^2P}{(2\pi)^2} e^{iP\cdot r}\frac{P^l}{P^2(K^2+c_0P^2)}
&=&
\int_0^\infty \frac{d|P|}{(2\pi)^2}\frac{1}{(K^2+c_0P^2)}\,\left[ 2\pi i \, \frac{r^l}{|r|} J_1\left(|P|\,|r|\right)\right] \nonumber \\
&=&
\frac{i}{2\pi}\frac{r^l}{r^2}\left[\frac{1}{K^2}- \frac{|r|}{\sqrt{c_0} |K|}
K_1\left(\frac{1}{\sqrt{c_0}}|K| \, |r|\right) \right] \, ,
\label{F_2_after_P}
\eeq
where $J_1(r)$ is the Bessel function of the first kind whereas $K_1(r)$ is the modified Bessel function of the second kind. By using the result in Eq. \eqref{F_2_after_P}, we can now consider the remaining $K$ integration in Eq. \eqref{F2_FT_with_ioffe} with the Ioffe time constraint, which reads
\bea
\label{F2_K_integration}
\frac{i}{2\pi}\frac{r^l}{r^2} \int_{K^2<\alpha}\frac{d^2K}{(2\pi)^2} \, e^{iK\cdot r'}  \left[\frac{K^m}{K^2}-\frac{|r|\, K^m}{\sqrt{c_0} |K|}
K_1\left(\frac{1}{\sqrt{c_0}}|K| \, |r|\right) \right].
\eea
Integrating the first term in the Eq.\eqref{F2_K_integration} over $K$ is straightforward and the result can be written as 
\bea
 \int_{K^2<\alpha}d^2K \, e^{iK\cdot r'} \frac{K^m}{K^2} = 2\pi i \, \frac{r'^m}{r'^2} \, \Big[1-J_0\left(\sqrt{\alpha} |r'|\right)\Big].
 \eea
The result of the $K$ Integration of the second term in Eq.\eqref{F2_K_integration} is more complicated and reads
\bea
&&\int_{K^2<\alpha}d^2K \, e^{iK\cdot r'} \frac{K^m}{|K|}K_1\left(\frac{1}{\sqrt{c_0}}|K| \, |r|\right)
 =
 2\pi i \, \frac{r'^m}{|r'|} \int_0^{\sqrt{\alpha}} d|K|\, |K|\, 
 \nonumber\\
 &&
 \hspace{9cm}
 \times \, 
 K_1\left(\frac{1}{\sqrt{c_0}}|K| \, |r|\right) J_1\left(|K|\, |r'|\right) \nonumber \\
&& =2\pi i \, \frac{r'^m}{|r'|}\,
 \frac{c_0\sqrt{c_0}}{|r|\left(c_0r'^2+r^2\right)}
 \\
 &&
 \times \, 
 \Bigg\{ |r'|+ \sqrt{\frac{\alpha}{{c_0}}}\,|r|\,|r'|\, J_2\Big(\sqrt{\alpha}|r'|\Big)
K_1\bigg(\sqrt{\frac{\alpha}{{c_0}}}|r|\bigg) 
-
\frac{\sqrt{\alpha}}{{c_0}}\,r^2\, J_1\Big(\sqrt{\alpha}|r'|\Big)
K_2\bigg(\sqrt{\frac{\alpha}{{c_0}}}|r|\bigg)
\Bigg\}.
\nonumber 
 \eea
After combining all the pieces we can write the result of Eq. \eqref{F2_FT_with_ioffe} with the Ioffe time constraint as 
\beq
\label{Result_KP_integration_with_Ioffe}
&&
\hspace{-1cm}
\int \frac{d^2P}{(2\pi)^2}\int_{K^2 < \alpha} \frac{d^2K}{(2\pi)^2} e^{iP\cdot r+iK\cdot r'}\frac{P^l}{P^2}\frac{K^m}{K^2+c_0P^2}=-\frac{1}{(2\pi)^2}\left(\frac{r'^m}{r'^2}\right) \frac{r^l}{r^2+c_0r'^2}
\nonumber \\
&&
\times
\bigg\{1-J_0\Big(\sqrt{\alpha}|r'|\Big)-c_0 \frac{r'^2}{r^2}\,J_0\Big(\sqrt{\alpha}|r'|\Big)
- 
\sqrt{{c_0}\,{\alpha}{}}\,  \frac{r'^2}{|r|} \, J_2\Big(\sqrt{\alpha}|r'|\Big)
K_1\left (\sqrt{\frac{\alpha}{{c_0}}}|r|\right)   \nonumber \\
&&
\hspace{6.5cm} +
{\sqrt{\alpha}}\,|r'|\, J_1\Big(\sqrt{\alpha}|r'|\Big)
K_2\left(\sqrt{\frac{\alpha}{{c_0}}}|r|\right)
\bigg\} \, .
\eeq
Now, by using Eq. \eqref{Result_KP_integration_with_Ioffe} we can simply write the splitting amplitude $F^{(2)}_{(\rmq\rmp-\rmq\rmg)}$ that corresponds to successive emission of a photon and a gluon with the Ioffe time restriction as  
\beq
&&
F^{(2)}_{(\rmq\rmp-\rmq\rmg)}\left[ (\rmp)[k_1^+,\omega-x_1]^{\lambda}; (\rmg)[k_2^+,\omega-x_2]^\eta; (\rmq)[p^+-k_1^+-k_2^+,\omega-x_3]_{ss''}\right]\nonumber\\
&&=\sum_{s'}\frac{1}{\sqrt{2\xi_1p^+}}\phi^{\lambda\bar\lambda}_{ss'}(\xi_1)\frac{1}{\sqrt{2\xi_2p^+}}\tilde{\phi}^{\eta\bar\eta}_{s's''}(\xi_1,\xi_2)\delta^{(2)}\left[\omega-\left(\xi_1x_1+\xi_2x_2+(1-\xi_1-\xi_2)x_3\right)\right]\nonumber\\
&&
\times
(-1)A^{\bar\eta}(x_3-x_2)\bigg[-\frac{1}{2\pi}\frac{(\omega-x_1)^\lambda}{\xi_1(\omega-x_1)^2+\xi_2(1-\xi_1-\xi_2)(x_3-x_2)^2}\bigg]\bar\xi_1
\nonumber\\
&&
\times
\bigg\{1-J_0\Big(\sqrt{\alpha}|x_3-x_2|\Big)-\frac{\xi_2(1-\xi_1-\xi_2)}{\xi_1}\frac{(x_3-x_2)^2}{(\omega-x_1)^2} J_0\Big(\sqrt{\alpha}|x_3-x_2|\Big)\nonumber\\
&&
-\sqrt{\alpha\frac{\xi_2(1-\xi_1-\xi_2)}{\xi_1}} \frac{(x_3-x_2)^2}{|\omega-x_1|}J_2\Big(\sqrt{\alpha}|x_3-x_2|\Big)
K_1\bigg(\sqrt{\frac{\alpha\xi_1}{\xi_2(1-\xi_1-\xi_2)}}|\omega-x_1|\bigg)\nonumber\\
&&
+\sqrt{\alpha}|x_3-x_2|J_1\Big(\sqrt{\alpha}|x_3-x_2| \Big)K_2\bigg(\sqrt{\frac{\alpha\xi_1}{\xi_2(1-\xi_1-\xi_2)}}|\omega-x_1|\bigg)\bigg\}.
\eeq
We can simplify this complicated expression. First of all, for future convenience, we can write the $\delta$-function as 
\beq
\label{simplification_1}
&&
\delta^{(2)}\left[ \omega -\left(\xi_1x_1+\xi_2x_2+(1-\xi_1-\xi_2)x_3\right)\right]=
\delta^{(2)}\left\{ \omega- \left[ \xi_1x_1+\bar\xi_1\left( \frac{\xi_2}{\bar\xi_1}x_2+\frac{\bar\xi_1-\xi_2}{\bar\xi_1}x_3\right)\right]\right\}\nonumber\\
&&
=\int_v \delta^{(2)}\left[ \omega-\left(\xi_1x_1+\bar\xi_1v\right)\right]\delta^{(2)}\left[v- \left\{\left(1-\frac{\xi_2}{\bar\xi_1}\right)x_3+\frac{\xi_2}{\bar\xi_1}x_2\right\}\right].
\eeq
Second, we can write the splitting amplitude $\tilde\phi$ in terms of $\phi$. The splitting amplitude $\tilde\phi^{\eta\bar\eta}_{s's''}$ is defined in Eq. \eqref{tildephi} as
\beq
\label{simplification_2}
\tilde{\phi}^{\eta\bar\eta}_{s's''}(\xi_1,\xi_2)&=&\frac{\xi_1}{\bar\xi_1} 
\left\{\left(2-2\xi_1-\xi_2\right)\delta^{\eta\bar\eta}\delta_{s's''}-i\epsilon^{\eta\bar\eta}\sigma^3_{s's''}\xi_2\right\}\nonumber\\
&=&
\xi_1
\left\{ \left(2-\frac{\xi_2}{\bar\xi_1}\right)\delta^{\eta\bar\eta}\delta_{s's''}-i\epsilon^{\eta\bar\eta}\sigma^3_{s's''}\frac{\xi_2}{\bar\xi_1}\right\}
=
\xi_1\phi^{\eta\bar\eta}_{s's''}\left(\frac{\xi_2}{\bar\xi_1}\right).
\eeq
Finally, using Eq. \eqref{simplification_1}, we can write
\beq
\label{simplification_3}
(\omega-x_1)^{\bar \lambda}=\bar\xi_1(v-x_1)^{\bar\lambda}.
\eeq
Thus, the splitting amplitude $F^{(2)}_{(\rmq\rmp-\rmq\rmg)}$ that corresponds to successive emission of a photon and a gluon with the Ioffe time restriction reads
\beq
&&
\label{F2_photon_gluon_final}
F^{(2)}_{(\rmq\rmp-\rmq\rmg)}\left[ (\rmp)[k_1^+,\omega-x_1]^{\lambda}; (\rmg)[k_2^+,\omega-x_2]^\eta; (\rmq)[p^+-k_1^+-k_2^+,\omega-x_3]_{ss''}\right]\nonumber\\
&&=\int_v
\delta^{(2)}\left[v- \left\{\left(1-\frac{\xi_2}{\bar\xi_1}\right)x_3+\frac{\xi_2}{\bar\xi_1}x_2\right\}\right]
\delta^{(2)}\left[ \omega-\left(\xi_1x_1+\bar\xi_1v\right)\right]\nonumber\\
&&
\times
\sum_{s'} 
\bigg[\frac{(-i)}{\sqrt{2\xi_1p^+}}\phi^{\lambda\bar\lambda}_{ss'}(\xi_1)\bigg]
\bigg[\frac{(-i)}{\sqrt{2\xi_2p^+}}{\phi}^{\eta\bar\eta}_{s's''}\bigg(\frac{\xi_2}{\bar\xi_1}\bigg)\bigg]
A^{\bar\eta}(x_3-x_2)
\bar{\cal A}_{\xi_2/\bar\xi_1}^{\bar\lambda}(v-x_1),
\eeq
where $A^{\bar\eta}(x_3-x_2)$ is the usual Weizs\''acker-Williams field in the quark-gluon splitting that is defined in Eq. \eqref{Ajxi}, while $\bar{\cal A}_{\xi_2/\bar\xi_1}^{\bar\lambda}(v-x_1)$ is the Weizs\''acker-Williams field-like term that appears in the splitting amplitude $F^{(2)}_{(\rmq\rmp-\rmq\rmg)}$ and it is defined as
\beq
\label{cal_A_first_ordering}
\bar{\cal A}^{\bar\lambda}_{\xi_2/\bar\xi_1}(v-x_1)=-\frac{1}{2\pi}\frac{\xi_1(v-x_1)^{\bar\lambda}}{\xi_1(v-x_1)^2+\frac{\xi_2}{\bar\xi_1}\left(1-\frac{\xi_2}{\bar\xi_1}\right)(x_3-x_2)^2}\bigg[1-f^{(\rmp\rmg)}_{\frac{\xi_2}{\bar\xi_1}}(\alpha;v-x_1,x_3-x_2 )\bigg],\nonumber\\
\eeq
where $f^{(\rmp\rmg)}_{\frac{\xi_2}{\bar\xi_1}}(\alpha;v-x_1,x_3-x_2 )$ is the correction due to the Ioffe time constraint which is defined as 
\beq
&&
f^{(\rmp\rmg)}_{\frac{\xi_2}{\bar\xi_1}}(\alpha;v-x_1,x_3-x_2 )=\Bigg\{ \bigg[1+\frac{1}{\xi_1}\frac{\xi_2}{\bar\xi_1}\bigg(1-\frac{\xi_2}{\bar\xi_1}\bigg)\frac{(x_3-x_2)^2}{(v-x_1)^2}\bigg]J_0\Big(\sqrt{\alpha}|x_3-x_2|\Big)\nonumber\\
&&
+\sqrt{\alpha\frac{1}{\xi_1}\frac{\xi_2}{\bar\xi_1}\bigg(1-\frac{\xi_2}{\bar\xi_1}\bigg)}\frac{(x_3-x_2)^2}{|v-x_1|}
J_2\Big(\sqrt{\alpha}|x_3-x_2|\Big)K_1\Bigg(\sqrt{\frac{\alpha\xi_1\bar\xi_1^2}{\xi_2\big(\bar\xi_1-\xi_2)}}|v-x_1|\Bigg)\nonumber\\
&&
-\sqrt{\alpha}|x_3-x_2|J_1\Big(\sqrt{\alpha}|x_3-x_2|\Big)K_2\Bigg(\sqrt{\frac{\alpha\xi_1\bar\xi_1^2}{\xi_2\big(\bar\xi_1-\xi_2)}}|v-x_1|\Bigg)\Bigg\},
\eeq
with $\alpha\equiv 2\frac{\xi_2}{\bar\xi_1}\bigg(1-\frac{\xi_2}{\bar\xi_1}\bigg)\frac{\bar\xi_1p^+}{\tau}$ being the upper bound of the momentum integral that is conjugate to quark-gluon dipole size, where $\tau$ can be identified as the longitudinal size of the target as in the case of quark-gluon splitting.


Finally, by using Eq. \eqref{F2_photon_gluon_final} we can write the first term in the quark-photon-gluon component of the dressed quark state (which defines the emission of the photon first and gluon later) as
\beq
\label{1st_ordering_qggamma}
&&
\hspace{-0.1cm}
g_sg_e\sum_{s''\lambda\eta}\int \frac{dk_1^+}{2\pi} \frac{dk_2^+}{2\pi} \frac{d^2k_1}{(2\pi)^2} \frac{d^2k_2}{(2\pi)^2} t^c_{\alpha\beta} \nonumber\\
&&
\hspace{1cm}
\times
F^{(2)}_{(\rmq\rmp-\rmq\rmg)}\left[ (\rmp)[k_1^+,k_1]^\lambda; (\rmg)[k_2^+,k_2]^c_\eta, (\rmq)[p^+-k_1^+-k_2^+, p-k_1-k_2]_{ss''} \right]
\nonumber\\
&&
\hspace{1cm}
\times
\left| (\rmq)[p^+-k_1^+-k_2^+,p-k_1-k_2]^\beta_{ss''}; (\rmg)[k^+_2,k_2]^c_{\eta}; (\rmp)[k_1^+,k_1]^{\lambda}\right\rangle_0
\nonumber\\
&&
\hspace{-0.2cm}
=
g_sg_e
\int \frac{dk_1^+}{2\pi} \frac{dk_2^+}{2\pi} t^c_{\alpha\beta}\int_{wvx_1x_2x_3}
\delta^{(2)}\left[v- \left\{\left(1-\frac{\xi_2}{\bar\xi_1}\right)x_3+\frac{\xi_2}{\bar\xi_1}x_2\right\}\right]
\delta^{(2)}\left[ \omega-\left(\xi_1x_1+\bar\xi_1v\right)\right]\
\nonumber\\
&&
\hspace{1cm}
\times
e^{-ip\cdot\omega}
\sum_{s'} 
\bigg[\frac{(-i)}{\sqrt{2\xi_1p^+}}\phi^{\lambda\bar\lambda}_{ss'}(\xi_1)\bigg]
\bigg[\frac{(-i)}{\sqrt{2\xi_2p^+}}{\phi}^{\eta\bar\eta}_{s's''}\bigg(\frac{\xi_2}{\bar\xi_1}\bigg)\bigg]
A^{\bar\eta}(x_3-x_2)
\bar{\cal A}_{\xi_2/\bar{\xi_1}}^{\bar\lambda}(v-x_1)\nonumber\\
&&
\hspace{1cm}
\times
\left| (\rmq)[p^+-k_1^+-k_2^+,x_3]^\beta_{ss''}; (\rmg)[k^+_2,x_2]^c_{\eta}; (\rmp)[k_1^+,x_1]^{\lambda}\right\rangle_0\,.
\eeq
If the Ioffe time restriction is neglected, then
\beq
\label{cal_A_without_Ioffe}
\bar{\cal A}^{\bar\lambda}_{\xi_2/{\bar\xi}_1}(v-x_1)\rightarrow -\frac{1}{2\pi}\frac{\xi_1(v-x_1)^{\bar\lambda}}{\xi_1(v-x_1)^2+\frac{\xi_2}{\bar\xi_1}\left(1-\frac{\xi_2}{\bar\xi_1}\right)(x_3-x_2)^2}\equiv { \cal A}^{\bar\lambda}_{\xi_2/\bar\xi_1}(v-x_1).
\eeq  

Let us now consider the second term in the quark-photon-gluon component of the dressed state which corresponds to emission of the gluon first and photon later. The calculation of this term can be performed exactly in a similar manner but the effect of the Ioffe time constraint is different between the two orderings of the photon and gluon emissions. 

The second ordering term in the mixed space reads
\beq
&&
\hspace{-1cm}
g_sg_e\sum_{s''\lambda\eta}\int \frac{dk_1^+}{2\pi} \frac{dk_2^+}{2\pi} \frac{d^2k_1}{(2\pi)^2} \frac{d^2k_2}{(2\pi)^2} t^c_{\alpha\beta} \nonumber\\
&&
\times
F^{(2)}_{(\rmq\rmg-\rmq\rmp)}\left[(\rmg)[k_2^+,k_2]^c_\eta; (\rmp)[k_1^+,k_1]^\lambda; (\rmq)[p^+-k_1^+-k_2^+, p-k_1-k_2]_{ss''} \right]
\nonumber\\
&&
\times
\left| (\rmq)[p^+-k_1^+-k_2^+,p-k_1-k_2]^\beta_{ss''}; (\rmg)[k^+_2,k_2]^c_{\eta}; (\rmp)[k_1^+,k_1]^{\lambda}\right\rangle_0
\nonumber\\
&&
\hspace{-1cm}
=
g_sg_e\sum_{s''\lambda\eta}\int \frac{dk_1^+}{2\pi} \frac{dk_2^+}{2\pi} \frac{d^2k_1}{(2\pi)^2} \frac{d^2k_2}{(2\pi)^2} t^c_{\alpha\beta}
\int_{x_iz_j}e^{-ik_1\cdot(z_1+x_1)-ik_2\cdot(z_2+x_2)-i(p-k_1-k_2)\cdot(z_3+x_3)}\nonumber\\
&&
\times
F^{(2)}_{(\rmq\rmg-\rmq\rmp)}\left[ (\rmg)[k_2^+,z_2]^c_\eta ; (\rmp)[k_1^+,z_1]^\lambda; (\rmq)[p^+-k_1^+-k_2^+, z_3]_{ss''} \right]
\nonumber\\
&&
\times
\left| (\rmq)[p^+-k_1^+-k_2^+,x_3]^\beta_{ss''}; (\rmg)[k^+_2,x_2]^c_{\eta}; (\rmp)[k_1^+,x_1]^{\lambda}\right\rangle_0\,.
\eeq
As for the first ordering term we can integrate over $k_1$ and $k_2$, use the resulting two $\delta$-functions to integrate over the variables $z_1$ and $z_2$, and finally perform a change of variables for $z_3\to \omega-x_3$ to write down the mixed space expression of the second term in the bare quark-photon-gluon component of the dressed quark state  as 
\beq
&&
g_sg_e\sum_{s''\lambda\eta}\int \frac{dk_1^+}{2\pi} \frac{dk_2^+}{2\pi} t^{c}_{\alpha\beta}\int_{wx_1x_2x_3}e^{-ip\cdot\omega}\nonumber\\
&&
\times
F^{(2)}_{(\rmq\rmg-\rmq\rmp)}\left[(\rmg)[k_2^+,\omega-x_2]^c_\eta; (\rmp)[k_1^+,\omega-x_1]^\lambda; (\rmq)[p^+-k_1^+-k_2^+, \omega-x_3]_{ss''} \right]\nonumber\\
&&
\times
\left| (\rmq)[p^+-k_1^+-k_2^+,x_3]^\beta_{ss''}; (\rmg)[k^+_2,x_2]^c_{\eta}; (\rmp)[k_1^+,x_1]^{\lambda}\right\rangle_0\,.
\eeq
Let us now calculate the Fourier transform of the splitting amplitude $F^{(2)}_{(\rmq\rmg-\rmq\rmp)}$. The explicit  expression of $F^{(2)}_{(\rmq\rmg-\rmq\rmp)}$ in full momentum space can be simply read off from Eq. \eqref{F2} by setting $\xi_1\leftrightarrow\xi_2$ and $k_1\leftrightarrow k_2$. Then the mixed space expression reads
\beq
&&
F^{(2)}_{(\rmq\rmg-\rmq\rmp)}\left[ (\rmg)[k_2^+,\omega-x_2]^\eta; (\rmp)[k_1^+,\omega-x_1]^{\lambda}; (\rmq)[p^+-k_1^+-k_2^+,\omega-x_3]_{ss''}\right]\nonumber\\
&&=\sum_{s'}
\frac{\phi^{\eta\bar\eta}_{ss'}(\xi_2)}{\sqrt{2\xi_1p^+}}
\frac{\tilde{\phi}^{\lambda\bar\lambda}_{s's''}(\xi_2,\xi_1)}{\sqrt{2\xi_2p^+}}
\int\frac{d^2p}{(2\pi)^2}\frac{d^2k_1}{(2\pi)^2}\frac{d^2k_2}{(2\pi)^2}e^{ik_1\cdot(w-x_1)+ik_2\cdot(w-x_2)+i(p-k_1-k_2)\cdot(\omega-x_3)}\nonumber\\
&&
\times
\frac{(\xi_2p-k_2)^{\bar\eta}}{(\xi_2p-k_2)^2}
\frac{ [\xi_1(p-k_2)-\bar\xi_2k_1]^{\bar\lambda} }{ \xi_1(\xi_2p-k_2)^2 + \xi_2(\xi_1p-k_1)^2 - (\xi_1k_2-\xi_2k_1)^2}\ .
\eeq
After performing the following change of variables:
\beq
\xi_2p-k_2&=&P,\\
\xi_1(p-k_2)-\bar\xi_2k_1&=&\bar\xi_2K,
\eeq 
we can integrate over $p$ to get the following expression:
\beq
&&
F^{(2)}_{(\rmq\rmg-\rmq\rmp)}\left[ (\rmg)[k_2^+,\omega-x_2]^\eta;(\rmp)[k_1^+,\omega-x_1]^{\lambda};  (\rmq)[p^+-k_1^+-k_2^+,\omega-x_3]_{ss''}\right]\nonumber\\
&&=\sum_{s'}
\frac{\phi^{\eta\bar\eta}_{ss'}(\xi_2)}{\sqrt{2\xi_2p^+}}
\frac{\tilde{\phi}^{\lambda\bar\lambda}_{s's''}(\xi_2,\xi_1)}{\sqrt{2\xi_1p^+}}
\delta^{(2)}\left[\omega-\left(\xi_1x_1+\xi_2x_2+(1-\xi_1-\xi_2)x_3\right)\right]\nonumber\\
&&
\times
\int \frac{d^2P}{(2\pi)^2}\frac{d^2K}{(2\pi)^2}e^{-iP\cdot\frac{(\omega-x_2)}{\bar \xi_2}-iK\cdot(x_3-x_1)}\frac{1}{\xi_2}\frac{P^{\bar\eta}}{P^2}\frac{K^{\bar\lambda}}{K^2+\frac{\xi_1(1-\xi_1-\xi_2)}{\xi_2{\bar\xi_2}^2}P^2}\ .
\eeq
For the integration over $P$ and $K$ the expression looks very similar to first ordering term, however  for this term the Ioffe time restriction is on momentum $P$ rather than momentum $K$, with 
\beq 
P^2 < 2\, \xi_2 \, {\bar{\xi_2}} \frac{p^+}{\tau} \equiv \beta.
\eeq
In our generic integral expression given in Eq.~\eqref{F2_FT_with_ioffe} we can first perform the integration over $K$ and the result reads
\bea 
\label{2nd_ordering_K_int}
\int\frac{d^2K}{(2\pi)^2} e^{iK\cdot r'}\frac{K^m}{K^2+c_0P^2}
= \frac{i}{2\pi} \, \sqrt{c_0}\,  \frac{r'^m}{|r'|}\,|P|\,
K_1\left(\sqrt{c_0}\,|P|\, |r'|\right).
\eea
The remaining integral over $P$ can be performed by using the result of Eq. \eqref{2nd_ordering_K_int} to get
\bea
&&\frac{i}{2\pi} \, \sqrt{c_0}\,  \frac{r'^m}{|r'|}\, 
\int_{P^2<\beta} \frac{d^2P}{(2\pi)^2}
 e^{iP\cdot r} \, \frac{P^l}{P^2} \,|P|\,
K_1\left(\sqrt{c_0}\,|P|\, |r'|\right) \nonumber \\
&&=
-\frac{\sqrt{c_0}}{(2\pi)^2}\, \frac{r'^m}{|r'|}\,\frac{r^l}{|r|}
\int_0^{\sqrt{\beta}} d |P| \, |P| \, K_1\left(\sqrt{c_0}\,|P|\, |r'|\right)
J_1\left(|P|\,|r|\right).
\eea

Thus, the final result after performing $K$ and $P$ integrations for the second ordering term reads 
\bea
\label{Result_PK_2nd_ordering_final}
&&
\int_{P^2<\beta} \frac{d^2P}{(2\pi)^2} \int\frac{d^2K}{(2\pi)^2} e^{iP\cdot r+iK\cdot r'}\frac{P^l}{P^2}\frac{K^m}{K^2+c_0P^2}=-\frac{1}{(2\pi)^2}\left(\frac{r'^m}{r'^2}\right) \frac{r^l}{r^2+c_0r'^2}  \\
&&
\times
\left[ 1+ \sqrt{\beta{{c_0}}}\,|r'|\, J_2\Big(\sqrt{\beta}|r|\Big)
K_1\Big(\sqrt{{\beta}{{c_0}}}|r'|\Big)
-
{\sqrt{\beta}}{{c_0}}\,\frac{r'^2}{|r|}\, J_1\Big(\sqrt{\beta}|r|\Big)
K_2\Big(\sqrt{{\beta}{{c_0}}}|r'|\Big)
\right].\nonumber
\eea
Finally, by using Eq.~\eqref{Result_PK_2nd_ordering_final} we can simply write the splitting amplitude $F^{(2)}_{(\rmq\rmg-\rmq\rmp)}$ that corresponds to successive emission of a gluon and a photon with the Ioffe time restriction as  
\beq
&&
F^{(2)}_{(\rmq\rmg-\rmq\rmp)}\left[(\rmg)[k_2^+,\omega-x_2]^\eta; (\rmp)[k_1^+,\omega-x_1]^{\lambda};  (\rmq)[p^+-k_1^+-k_2^+,\omega-x_3]_{ss''}\right]\\
&&=\sum_{s'}\frac{1}{\sqrt{2\xi_2p^+}}\phi^{\eta\bar\eta}_{ss'}(\xi_2)\frac{1}{\sqrt{2\xi_1p^+}}\tilde{\phi}^{\lambda\bar\lambda}_{s's''}(\xi_2,\xi_1)\delta^{(2)}\left[\omega-\left(\xi_1x_1+\xi_2x_2+(1-\xi_1-\xi_2)x_3\right)\right]\nonumber\\
&&
\times
(-1)A^{\bar\lambda}(x_3-x_1)\bigg[-\frac{1}{2\pi}\frac{(\omega-x_2)^{\bar\eta}}{\xi_2(\omega-x_2)^2+\xi_1(1-\xi_1-\xi_2)(x_3-x_1)^2}\bigg]\bar\xi_2
\nonumber\\
&&
\times
\bigg\{1+\sqrt{\beta\frac{\xi_1(1-\xi_1-\xi_2)}{\xi_2\bar\xi_2^2}}|x_3-x_1|
J_2\bigg(\sqrt{\beta}\frac{|\omega-x_2|}{\bar\xi_2}\bigg)
K_1\bigg(\sqrt{\beta\frac{\xi_1(1-\xi_1-\xi_2)}{\xi_2\bar\xi_2^2}}|x_3-x_1|\bigg)
\nonumber\\
&&
-\sqrt{\beta}\frac{\xi_1(1-\xi_1-\xi_2)}{\xi_2\bar\xi_2}\frac{(x_3-x_1)^2}{|\omega-x_2|}
J_1\bigg(\sqrt{\beta}\frac{|\omega-x_2|}{\bar\xi_2}\bigg)
K_2\bigg(\sqrt{\beta\frac{\xi_1(1-\xi_1-\xi_2)}{\xi_2\bar\xi_2^2}}|x_3-x_1|\bigg)\bigg\}.\nonumber
\eeq

We can simplify the above expression by using the analogues of Eqs. \eqref{simplification_1}-\eqref{simplification_3}. First, we rewrite the $\delta$-function as 
\beq
\label{simplification_1_1}
&&
\delta^{(2)}\left[ \omega -\left(\xi_1x_1+\xi_2x_2+(1-\xi_1-\xi_2)x_3\right)\right]=
\delta^{(2)}\left\{ \omega- \left[ \xi_2x_2+\bar\xi_2\left( \frac{\xi_1}{\bar\xi_2}x_1+\frac{\bar\xi_2-\xi_1}{\bar\xi_2}x_3\right)\right]\right\}\nonumber\\
&&
= \int_v\delta^{(2)}\left[ \omega-\left(\xi_2x_2+\bar\xi_2v\right)\right]\delta^{(2)}\left[v- \left\{\left(1-\frac{\xi_1}{\bar\xi_2}\right)x_3+\frac{\xi_1}{\bar\xi_2}x_1\right\}\right].
\eeq
Then, we rewrite the splitting amplitude $\tilde\phi$ in terms of $\phi$ as
\beq
\label{simplification_2_1}
\tilde{\phi}^{\lambda\bar\lambda}_{s's''}(\xi_2,\xi_1)&=&\frac{\xi_2}{\bar\xi_2} 
\left\{\left(2-2\xi_2-\xi_1\right)\delta^{\lambda\bar\lambda}\delta_{s's''}-i\epsilon^{\lambda\bar\lambda}\sigma^3_{s's''}\xi_1\right\}\nonumber\\
&=&
\xi_2
\left\{ \left(2-\frac{\xi_1}{\bar\xi_2}\right)\delta^{\lambda\bar\lambda}\delta_{s's''}-i\epsilon^{\lambda\bar\lambda}\sigma^3_{s's''}\frac{\xi_1}{\bar\xi_2}\right\}
=
\xi_2\phi^{\lambda\bar\lambda}_{s's''}\left(\frac{\xi_1}{\bar\xi_2}\right).
\eeq
Finally, using Eq.~\eqref{simplification_1_1}, we can write
\beq
(\omega-x_2)^{\bar \lambda}=\bar\xi_2(v-x_2)^{\bar\lambda}.
\eeq
Eventually, the splitting amplitude $F^{(2)}_{(\rmq\rmg-\rmq\rmp)}$ that corresponds to successive emission of a gluon and a photon with the Ioffe time restriction reads
\beq
&&
\label{F2_gluon_photon_final}
F^{(2)}_{(\rmq\rmg-\rmq\rmp)}\left[ (\rmg)[k_2^+,\omega-x_2]^\eta;(\rmp)[k_1^+,\omega-x_1]^{\lambda};  (\rmq)[p^+-k_1^+-k_2^+,\omega-x_3]_{ss''}\right]\nonumber\\
&&=\int_v
\delta^{(2)}\left[v- \left\{\left(1-\frac{\xi_1}{\bar\xi_2}\right)x_3+\frac{\xi_1}{\bar\xi_2}x_1\right\}\right]
\delta^{(2)}\left[ \omega-\left(\xi_2x_2+\bar\xi_2v\right)\right]\nonumber\\
&&
\times
\sum_{s'} 
\bigg[\frac{(-i)}{\sqrt{2\xi_2p^+}}\phi^{\eta\bar\eta}_{ss'}(\xi_2)\bigg]
\bigg[\frac{(-i)}{\sqrt{2\xi_1p^+}}{\phi}^{\lambda\bar\lambda}_{s's''}\bigg(\frac{\xi_1}{\bar\xi_2}\bigg)\bigg]
A^{\bar\lambda}(x_3-x_1)
\bar{\cal A}^{\bar\eta}_{\xi_1/\bar\xi_2}(v-x_2),
\eeq
where $A^{\bar\lambda}(x_3-x_1)$ is the usual Weizs\''acker-Williams field in the quark-photon splitting that is defined in Eq. \eqref{Ajxi} and $\bar{\cal A}_{\xi_1/\bar\xi_2}^{\bar\eta}(v-x_2)$ is the Weizs\''acker-Williams field-like term that appears in the splitting amplitude $F^{(2)}_{(\rmq\rmg-\rmq\rmp)}$ and it is defined as 
\beq
\label{cal_A_second_ordering}
\bar{\cal A}^{\bar\eta}_{\xi_1/\bar\xi_2}(v-x_2)=\frac{-1}{2\pi}\frac{\xi_2(v-x_2)^{\bar\eta}}{\xi_2(v-x_2)^2+\frac{\xi_1}{\bar\xi_2}\left(1-\frac{\xi_1}{\bar\xi_2}\right)(x_3-x_1)^2}
\bigg[1-f^{(\rmg\rmp)}_{\frac{\xi_1}{\bar\xi_2}}\left(\beta;v-x_2,x_3-x_1\right)\bigg],\nonumber\\
\eeq
where $f^{(\rmg\rmp)}_{\frac{\xi_1}{\bar\xi_2}}\left(\beta;v-x_2,x_3-x_1\right)$ is the correction due to the Ioffe time constraint  which can be written as 
\beq
&&
f^{(\rmg\rmp)}_{\frac{\xi_1}{\bar\xi_2}}(\beta; v-x_2,x_3-x_1)=\\
&&
=
\sqrt{\beta}\frac{1}{\xi_2}\frac{\xi_1}{\bar\xi_2}\bigg(1-\frac{\xi_1}{\bar\xi_2}\bigg)\frac{(x_3-x_1)^2}{|v-x_2|}J_1\Big(\sqrt{\beta}|v-x_2|\Big)K_2\bigg(\sqrt{\beta\frac{1}{\xi_2}\frac{\xi_1}{\bar\xi_2}\bigg(1-\frac{\xi_1}{\bar\xi_2}\bigg)}|x_3-x_1|\bigg)\nonumber\\
&&
-\sqrt{\beta\frac{1}{\xi_2}\frac{\xi_1}{\bar\xi_2}\bigg(1-\frac{\xi_1}{\bar\xi_2}\bigg)}|x_3-x_1|J_2\Big(\sqrt{\beta}|v-x_2|\Big)
K_1\bigg(\sqrt{\beta\frac{1}{\xi_2}\frac{\xi_1}{\bar\xi_2}\bigg(1-\frac{\xi_1}{\bar\xi_2}\bigg)}|x_3-x_1|\bigg),\nonumber
\eeq
with $\beta\equiv 2\xi_2\bar\xi_2\frac{p^+}{\tau}$ that is the upper bound of the momentum integration which is conjugate to the quark-gluon dipole size. As in the case of the first ordering term, when the Ioffe time is neglected we get
\beq
\bar{\cal A}^{\bar\eta}_{\xi_1/\bar\xi_2}(v-x_2)\rightarrow\frac{-1}{2\pi}\frac{\xi_2(v-x_2)^{\bar\eta}}{\xi_2(v-x_2)^2+\frac{\xi_1}{\bar\xi_2}\left(1-\frac{\xi_1}{\bar\xi_2}\right)(x_3-x_1)^2}\equiv{\cal A}^{\bar\eta}_{\xi_1/\bar\xi_2}(v-x_2).
\eeq

Note again, that the corrections due to the Ioffe time constraint are different between the two emission orderings in the quark-photon-gluon component of the dressed quark state. This is due to the fact that the two orderings have different momentum that is restricted by the Ioffe time constraint, which changes the result of the momentum integrations performed in order to  Fourier transform the splitting amplitudes $F^{(2)}_{\rmq\rmp -\rmq\rmg}$ and $F^{(2)}_{\rmq\rmg -\rmq\rmp}$.

By using Eq. \eqref{F2_gluon_photon_final} we can write the second term in the quark-photon-gluon component of the dressed quark state (which defines the emission of the gluon first and photon later) as
\beq
\label{2nd_ordering_qggamma}
&&
\hspace{-0.1cm}
g_sg_e\sum_{s''\lambda\eta}\int \frac{dk_1^+}{2\pi} \frac{dk_2^+}{2\pi} \frac{d^2k_1}{(2\pi)^2} \frac{d^2k_2}{(2\pi)^2} t^c_{\alpha\beta} \nonumber\\
&&
\hspace{1cm}
\times
F^{(2)}_{(\rmq\rmg-\rmq\rmp)}\left[  (\rmg)[k_2^+,k_2]^c_\eta; (\rmp)[k_1^+,k_1]^\lambda; (\rmq)[p^+-k_1^+-k_2^+, p-k_1-k_2]_{ss''} \right]
\nonumber\\
&&
\hspace{1cm}
\times
\left| (\rmq)[p^+-k_1^+-k_2^+,p-k_1-k_2]^\beta_{ss''}; (\rmg)[k^+_2,k_2]^c_{\eta}; (\rmp)[k_1^+,k_1]^{\lambda}\right\rangle_0
\nonumber\\
&&
\hspace{-0.2cm}
=
g_sg_e
\int \frac{dk_1^+}{2\pi} \frac{dk_2^+}{2\pi} t^c_{\alpha\beta}\int_{wvx_1x_2x_3}
\delta^{(2)}\left[v- \left\{\left(1-\frac{\xi_1}{\bar\xi_2}\right)x_3+\frac{\xi_1}{\bar\xi_2}x_1\right\}\right]
\delta^{(2)}\left[ \omega-\left(\xi_2x_2+\bar\xi_2v\right)\right]\
\nonumber\\
&&
\hspace{1cm}
\times
e^{-ip\cdot\omega}
\sum_{s'} 
\bigg[\frac{(-i)}{\sqrt{2\xi_2p^+}}\phi^{\eta\bar\eta}_{ss'}(\xi_2)\bigg]
\bigg[\frac{(-i)}{\sqrt{2\xi_1p^+}}{\phi}^{\lambda\bar\lambda}_{s's''}\bigg(\frac{\xi_1}{\bar\xi_2}\bigg)\bigg]
A^{\bar\lambda}(x_3-x_1)
\bar{\cal A}^{\bar\eta}_{\xi_1/\bar\xi_2}(v-x_2)\nonumber\\
&&
\hspace{1cm}
\times
\left| (\rmq)[p^+-k_1^+-k_2^+,x_3]^\beta_{ss''}; (\rmg)[k^+_2,x_2]^c_{\eta}; (\rmp)[k_1^+,x_1]^{\lambda}\right\rangle_0\,.
\eeq
By combining Eqs. \eqref{trivial}, \eqref{qgamma_component}, \eqref{qg_component}, \eqref{1st_ordering_qggamma} and \eqref{2nd_ordering_qggamma} together and setting $p\to0$, we get Eq. \eqref{Dressed_mixed_space}.
%
\subsection{Outgoing wave function}
\label{Section:Outgoing_WaveFunction}

The dressed quark state scatters through the target eikonally, which means that each bare state component of the dressed quark state rotates in the color space by picking up an $S$-matrix. Thus, the outgoing wave function - when written in terms of the bare states - reads
\beq
\label{OutWF_1}
&&
\left| (\rmq)[p^+,0]_s^{\alpha}\right\rangle_{\rm out}= \int_\omega S_F^{\alpha\beta}(\omega) \left|(\rmq)[p^+,\omega]_s^{\beta}\right\rangle_0 \nonumber\\
&&
+g_e\sum_{s'\lambda}\int\frac{dk_1^+}{2\pi}\int_{\omega v x_1} S_F^{\alpha\beta}(v)
\bigg[\frac{(-i)}{\sqrt{2\xi_1p^+}}\phi^{\lambda\bar\lambda}_{ss'}(\xi_1)\bigg]
A^{\bar\lambda}(v-x_1)\delta^{(2)}\left[\omega-(\bar\xi_1v+\xi_1x_1)\right]\nonumber\\
&&
\hspace{7cm}
\times
\left| (\rmq)[p^+-k_1^+,v]^\beta_{s}; (\rmp)[k_1^+,x_1]^\lambda\right\rangle_0\nonumber\\
&&
+g_s\sum_{s'\eta}\int\frac{dk_2^+}{2\pi}\int_{\omega v x_2}t^c_{\alpha\beta}S_F^{\beta\sigma}(v)S^{cd}_A(x_2) 
\bigg[\frac{(-i)}{\sqrt{2\xi_2p^+}}\phi^{\eta\bar\eta}_{ss'}(\xi_2)\bigg]
\bar{A}^{\bar\eta}(v-x_2)\delta^{(2)}\left[ \omega-(\bar\xi_2v+\xi_2x_2)\right]\nonumber\\
&&
\hspace{7cm}
\times
\left| (\rmq)[p^+-k_2^+,v]^\sigma_{s'}; (\rmg)[k_2^+,x_2]^c_{\eta}\right\rangle_0\nonumber\\
&&
+g_sg_e \sum_{s's''}\sum_{\lambda\eta}\int \frac{dk_1^+}{2\pi}\frac{dk_2^+}{2\pi} \int_{wvx_1x_2x_3} 
t^c_{\alpha\beta}S_F^{\beta\sigma}(x_3)S_A^{cd}(x_2)
\nonumber\\
&&
\times
\Bigg\{ 
\delta^{(2)}\left[v-\left\{\left(1-\frac{\xi_2}{\bar\xi_1}\right)x_3+\frac{\xi_2}{\bar\xi_1}x_2\right\}\right]
\delta^{(2)}\Big[\omega-\left(\xi_1x_1+\bar\xi_1v\right)\Big]
\nonumber\\
&&
\hspace{0.7cm}
\times
\bigg[\frac{(-i)}{\sqrt{2\xi_1p^+}}\phi^{\lambda\bar\lambda}_{ss'}(\xi_1)\bigg]
\bigg[\frac{(-i)}{\sqrt{2\xi_2p^+}}\phi^{\eta\bar\eta}_{s's''}\left(\frac{\xi_2}{\bar\xi_1}\right)\bigg]
A^{\bar\eta}(x_3-x_2) \bar{\cal A}^{\bar\lambda}_{\xi_2/\bar\xi_1}(v-x_1)
\\
&&
\hspace{0.5cm}
+\, 
\delta^{(2)}\left[v-\left\{\left(1-\frac{\xi_1}{\bar\xi_2}\right)x_3+\frac{\xi_1}{\bar\xi_2}x_1\right\}\right]
\delta^{(2)}\Big[\omega-\left(\xi_2x_2+\bar\xi_2v\right)\Big]
\nonumber\\
&&
\hspace{0.7cm}
\times
\bigg[\frac{(-i)}{\sqrt{2\xi_2p^+}}\phi^{\eta\bar\eta}_{ss'}(\xi_2)\bigg]
\bigg[\frac{(-i)}{\sqrt{2\xi_1p^+}}\phi^{\lambda\bar\lambda}_{s's''}\left(\frac{\xi_1}{\bar\xi_2}\right)\bigg]
A^{\bar\lambda}(x_3-x_1) \bar{\cal A}^{\bar\eta}_{\xi_1/\bar\xi_2}(v-x_2)\Bigg\}
\nonumber\\
&&
\hspace{5cm}
\times
\left| (\rmq)[p^+-k_1^+-k_2^+,x_3]^{\sigma}_{s''}, (\rmg)[k_2^+,x_2]^d_{\eta}, (\rmp) [k_1^+,x_1]^{\lambda}\right\rangle_0\,.\nonumber
\eeq
We would like to write the outgoing wave function in terms of dressed components instead of bare components. In order to do so, one should realize that the dressed states are written as decomposition of various bare states. Let us first explain how to rewrite outgoing wave function in terms of dressed states in a schematic way. The dressed states can be written in terms of the bare ones up to ${\cal O}(g_eg_s)$ as 
\beq
\label{D_ito_B}
|q\rangle_D&\simeq& |q\rangle_0+g_eF^{(1)}_{(q\gamma)}|q\gamma\rangle_0+g_sF^{(1)}_{(qg)}|qg\rangle_0+g_eg_s\left[F^{(2)}_{(q\gamma g)}+F^{(2)}_{(qg\gamma)}\right]|q\gamma g\rangle_0\,, \nonumber\\
|q\gamma\rangle_D &\simeq& |q\gamma\rangle_0+g_sF^{(1)}_{(qg)}|q\gamma g\rangle_0\,,\\
|qg\rangle_D &\simeq& |qg\rangle_0+g_eF^{(1)}_{(q\gamma)}|q\gamma g\rangle_0\,,\nonumber\\
|q\gamma g\rangle_D &\simeq& |q\gamma g\rangle_0\,.\nonumber
\eeq
The outgoing wave function, when written schematically in terms of the bare states (an analogue of Eq. \eqref{OutWF_1}), reads
\beq
\label{out_sch_1}
|q\rangle_{out}\simeq &&
S_F(\omega)|q\rangle_0+g_eF^{(1)}_{(q\gamma)}S_F(v)|q\gamma\rangle_0
+g_sF^{(1)}_{(qg)}S_F(v)S_A(x_2) |qg\rangle_0\nonumber\\
&&
+g_eg_s\left[F^{(2)}_{(q\gamma g)}+F^{(2)}_{(qg\gamma)}\right] S_F(x_3)S_A(x_2)  |q\gamma g\rangle_0\,.
\eeq
We can now rewrite Eq. \eqref{OutWF_1} in terms of the dressed components by using Eq. \eqref{D_ito_B} and group the dressed components. Finally, the outgoing wave function when written schematically in terms of the dressed components reads
\beq
\label{out_sch_2}
&&
|q\rangle_{out}= S_F(\omega)|q\rangle_D+g_eF^{(1)}_{(q\gamma)}\left[ S_F(v)-S_F(\omega)\right]|q\gamma\rangle_D
+g_sF^{(1)}_{(qg)}\left[ S_F(v)S_A(x_2)-S_F(\omega)\right] |qg\rangle_D \nonumber\\
&&
+g_eg_s\bigg\{ \Big[ F^{(2)}_{(q\gamma g)}\Big( S_F(x_3)S_A(x_2)-S_F(\omega)\Big)-F^{(1)}_{(q\gamma)}F^{(1)}_{(qg)}\Big( S_F(v)-S_F(\omega)\Big)\Big]\\
&&
\hspace{1cm}
+ 
\Big[ F^{(2)}_{(qg\gamma)}\Big( S_F(x_3)S_A(x_2)-S_F(\omega)\Big)-F^{(1)}_{(qg)}F^{(1)}_{(q\gamma)}\Big( S_F(v)S_A(x_3)-S_F(\omega)\Big)\Big]\bigg\} |q\gamma g\rangle_D\,.\nonumber
\eeq
Finally, with the guidance of the schematic expression, Eq.\eqref{out_sch_2}, we can write the outgoing wave function that was given in Eq. \eqref{OutWF_1}  in terms of the dressed states as in Eq. \eqref{out_full}.
%


\section{Expansion in the back-to-back correlation limit}
\label{Section:AppendixB}

In this appendix we provide the details of the expansion in the back-to-back correlation limit where the small parameters are the quark-gluon dipole sizes in the amplitude and in the complex conjugate amplitude ($r$ and $\bar r$). Thus, the Taylor expansion of a dipole $s\Big(b+\frac{r}{2},\bar b+\frac{\bar r}{2}\Big)$ simply reads
\beq
\label{example_expansion}
&&
s\Big(b+\frac{r}{2},\bar b+\frac{\bar r}{2}\Big)=\frac{1}{N_c}\tr\bigg[S_F\Big(b+\frac{r}{2}\Big)S_F^\dagger\Big(\bar b+\frac{\bar r}{2}\Big)\bigg]\\
&&
=\frac{1}{N_c}\tr\Bigg\{\bigg[ S_F(b)+\frac{1}{2}r^i\partial^iS_F(b)+\frac{1}{8}r^ir^j\partial^i\partial^jS_F(b)\bigg]
\bigg[S^\dagger_F({\bar b})+\frac{1}{2}\bar r^k\partial^kS_F^\dagger({\bar b})+\frac{1}{8}\bar r^k\bar r^l\partial^k\partial^lS_F^\dagger({\bar b})\bigg]\Bigg\}.\nonumber
\eeq

We start with the expression that we get for the production cross section, which is written in terms of the new variables $r,\bar r, b,\bar b, \gamma$ and $\bar \gamma$, Eq.~\eqref{X_section_in_r_b}, and use Eq.~\eqref{example_expansion} to expand each term separately. \\

\noindent (i) {\it ${\cal A}_{\xi_2}^\lambda(-\bar\gamma){\cal A}_{\xi_2}^\lambda(-\gamma)$ term}: When expanded in powers of $r$ and $\bar r$ 
\beq
\label{expansion_Q}
Q\left(\bar b-\frac{\bar r}{2},b-\frac{r}{2}, b+\frac{r}{2}, \bar b+\frac{\bar r}{2}\right)&=&1-r^i\bar r^j\frac{1}{N_c}\tr\left( \partial^iS_F(b)S_F^\dagger({\bar b})\partial^jS_F({\bar b})S_F^\dagger(b)\right)\nonumber\\
&&
\hspace{-3cm}
-\frac{1}{2}r^ir^j\frac{1}{N_c}\tr\left(\partial^iS_F(b)\partial^jS^\dagger_F(b)\right)
-\frac{1}{2}{\bar r}^i{\bar r}^j\frac{1}{N_c}\tr\left(\partial^iS_F({\bar b})\partial^jS^\dagger_F({\bar b})\right)
,\nonumber\\
s\left(b+\frac{r}{2},b-\frac{r}{2}\right)&=&1-\frac{1}{2}r^ir^j\frac{1}{N_c}\tr\left(\partial^iS_F(b)\partial^jS^\dagger_F(b)\right),\\
s\left(\bar b-\frac{\bar r}{2},\bar b+\frac{\bar r}{2}\right)&=&1-\frac{1}{2}{\bar r}^i{\bar r}^j\frac{1}{N_c}\tr\left(\partial^iS_F({\bar b})\partial^jS^\dagger_F({\bar b})\right).\nonumber
\eeq
Using Eq.~\eqref{expansion_Q}, the ${\cal O}(N_c^2)$ contribution to the operator structure of ${\cal A}_{\xi_2}^\lambda(-\bar\gamma){\cal A}_{\xi_2}^\lambda(-\gamma)$ term can be written as
\beq
\label{tildeAtildeA_Nc}
&&
\hspace{-0.4cm}
\frac{N_c^2}{2}\frac{1}{N_c}
\tr\Bigg\{ \! \bigg[ \!\Big(S_F(b)\!-\!S_F({b+\xi_1\gamma})\Big) \! + \!  \frac{1}{2}r^i\partial^i\Big(S_F(b) \!- \!(1 \!- \!2\xi_2)S_F({b+\xi_1\gamma})\Big)\nonumber\\
&&
\hspace{2cm}
+\frac{1}{8}r^ir^j\partial^i\partial^j\Big(S_F(b)\!-\!(1\!-\!2\xi_2)^2S_F({b+\xi_1\gamma})\Big)\bigg]\nonumber\\
&&
\hspace{-0.2cm}
\times\bigg[
\!\Big(S_F^\dagger({\bar b})\!-\!S_F^\dagger({\bar b+\xi_1\bar \gamma})\Big)
 \! + \!  \frac{1}{2}\bar r^k\partial^k\Big(S_F^\dagger({\bar b}) \!- \!(1 \!- \!2\xi_2)S^\dagger_F({\bar b+\xi_1\bar \gamma})\Big)\nonumber\\
 &&
 \hspace{2cm}
+ \frac{1}{8}\bar r^k\bar r^l\partial^k\partial^l\Big(S_F^\dagger({\bar b})-(1\!-\!2\xi_2)^2S_F^\dagger({\bar b+\xi_1\bar \gamma})\Big)\bigg]\Bigg\}\nonumber\\
&&
-\frac{N_c^2}{2}r^i{\bar r}^j\frac{1}{N_c}\tr\left(\partial^iS_F(b)S_F^\dagger({\bar b})\partial^jS_F({\bar b})S_F^\dagger(b)\right)s(b,\bar b)\nonumber\\
&&
-\frac{N_c^2}{2}r^ir^j\frac{1}{N_c}\tr\left(\partial^iS_F(b)\partial^jS^\dagger_F(b)\right)\left[ s(b,\bar b)-s(b,\bar b+\xi_1\bar\gamma)\right]\nonumber\\
&&
-\frac{N_c^2}{2}{\bar r}^i{\bar r}^j\frac{1}{N_c}\tr\left(\partial^iS_F({\bar b})\partial^jS^\dagger_F({\bar b})\right)\left[s(b,{\bar b})-s(b+\xi_1\gamma,\bar b)\right]\, .
\eeq
One can also calculate in a similar way the ${\cal O}(1)$ contribution to the operator structure of ${\cal A}_{\xi_2}^\lambda(-\bar\gamma){\cal A}_{\xi_2}^\lambda(-\gamma)$ term as
\beq
\label{tildeAtildeA_1}
&&
\hspace{-0.2cm}
-\frac{1}{2}\frac{1}{N_c}\tr \Bigg\{ \!\bigg[ \! \Big(S_F(b)-S_F({b+\xi_1\gamma})\Big) \!-\! \frac{1}{2}r^i\partial^i\Big(S_F(b)\!+\!(1\!-\!2\xi_2)S_F({b+\xi_1\gamma})\Big) \nonumber\\
&&
\hspace{2cm}
+  \frac{1}{8}r^ir^j\partial^i\partial^j\Big(S_F(b)-(1-2\xi_2)^2S_F({b+\xi_1\gamma})\Big)\! \bigg]\nonumber\\
&&
\hspace{1.3cm}
\times
\bigg[ \!\Big( S^\dagger_F({\bar b})-S^{\dagger}_F({\bar b+\xi_1\bar\gamma})\Big)-\frac{1}{2}{\bar r}^k\partial^k\Big(S^\dagger_F({\bar b})+(1-2\xi_2)S^\dagger_F({\bar b+\xi_1\bar{\gamma}})\Big)
\nonumber\\
&&
\hspace{2cm}+\frac{1}{8}{ \bar r}^k{ \bar r}^l\partial^k
\partial^k\Big(S^\dagger_F({\bar b})-(1-2\xi_2)^2S^\dagger_F({\bar b+\xi_1\bar\gamma})\Big)\!\bigg] \!\Bigg\}.
\eeq
Now let us consider the expansion of ${\cal A}_{\xi_2}^\lambda(-\gamma)$ in the small $r$ limit:
\beq
\label{expansiontildeA}
{\cal A}_{\xi_2}^\lambda(-\gamma)=-\frac{1}{2\pi}\frac{\xi_1(-\gamma)^\lambda}{\xi_1\gamma^2+\xi_2\bar{\xi_2}r^2}
=-\frac{1}{2\pi}\frac{-\gamma^\lambda}{\gamma^2}\frac{1}{1+\frac{\xi_2\bar\xi_2}{\xi_1}\frac{r^2}{\gamma^2}}
\simeq A^\lambda(-\gamma)\bigg(1-\frac{\xi_2\bar\xi_2}{\xi_1}\frac{r^2}{\gamma^2}\bigg).
\eeq
Using Eqs.~\eqref{tildeAtildeA_Nc}, \eqref{tildeAtildeA_1} and \eqref{expansiontildeA}, we can write the expanded expression of  
${\cal A}_{\xi_2}^\lambda(-\bar\gamma){\cal A}_{\xi_2}^\lambda(-\gamma)$  up to ${\cal O}(r^3)$ as
\beq
\label{tildeAtildeA_full}
&&
\hspace{0cm}
A^\lambda(-\bar\gamma)A^{\lambda}(-\gamma)\frac{1}{N_c}\tr\Bigg\{
\frac{N_c^2-1}{2}\bigg[1-\frac{\xi_2\bar\xi_2}{\xi_1}\bigg(\frac{r^2}{\gamma^2}+\frac{\bar r^2}{\bar \gamma^2}\bigg)\bigg] 
\nonumber\\
&&
\hspace{4cm}
\times
\Big(S_F(b)-S_F({b+\xi_1\gamma})\Big)\Big(S^\dagger_F({\bar b})-S^\dagger_F({\bar b+\xi_1\bar \gamma})\Big)\nonumber\\
&&
\hspace{0.5cm}
+\frac{1}{2}r^i\bigg[\frac{N_c^2+1}{2}\partial^iS_F(b)-\frac{N_c^2-1}{2}(1-2\xi_2)\partial^iS_F({b+\xi_1\gamma})\bigg]\Big(S^\dagger_F({\bar b})-S^\dagger_F({\bar b+\xi_1\bar\gamma})\Big)\nonumber\\
&&
\hspace{0.5cm}
+\frac{1}{2}\bar r^i\Big(S_F(b)-S_F({b+\xi_1\gamma})\Big)\bigg[\frac{N_c^2+1}{2}\partial^iS^\dagger_F({\bar b})-\frac{N_c^2-1}{2}(1-2\xi_2)\partial^iS^\dagger_F({\bar b+\xi_1\bar\gamma})\bigg]\nonumber\\
&&
\hspace{0.5cm}
+\frac{N_c^2-1}{2}\frac{1}{8}r^ir^j\bigg[\partial^i\partial^jS_F(b)-(1-2\xi_2)^2\partial^i\partial^jS^{\dagger}_F({b+\xi_1\gamma})\bigg]\Big(S^\dagger_F({\bar b})-S^\dagger_F({\bar b+\xi_1\bar\gamma})\Big)\nonumber\\
&&
\hspace{0.5cm}
+\frac{N_c^2-1}{2}\frac{1}{8}\bar r^i\bar r^j \Big(S_F(b)-S_F({b+\xi_1\gamma})\Big)\bigg[\partial^i\partial^jS^\dagger_F({\bar b})-(1-\xi_2)^2\partial^i\partial^jS^\dagger_F({\bar b+\xi_1\bar\gamma})\bigg]\nonumber\\
&&
\hspace{0.5cm}
+\frac{N_c^2-1}{2}\frac{1}{4}r^i\bar r^j\bigg[ \partial^iS_F(b)\partial^jS^\dagger_F({\bar b})+(1-2\xi_2)^2\partial^iS_F({b+\xi_1\gamma})\bigg]\nonumber\\
&&
\hspace{0.5cm}
-\frac{N_c^2+1}{2}\frac{1}{4}r^i\bar r^j(1-2\xi_2)\bigg[\partial^iS_F(b)\partial^jS^\dagger_F({\bar b+\xi_1\bar\gamma})+\partial^iS_F({b+\xi_1\gamma})\partial^jS^\dagger_F({\bar b})\bigg]\Bigg\}\nonumber\\
&&
\hspace{0cm}
-
A^\lambda(-\bar\gamma)A^{\lambda}(-\gamma) \frac{N_c^2}{2}r^i\bar r^j\frac{1}{N_c}\tr\Big(\partial^iS_F(b)S^\dagger_F({\bar b})\partial^jS_F({\bar b})S^\dagger_F(b)\Big)s(b,\bar b)\nonumber\\
&&
-A^\lambda(-\bar\gamma)A^{\lambda}(-\gamma) \frac{N_c^2}{2}r^ir^j\frac{1}{N_c}\tr\left(\partial^iS_F(b)\partial^jS^\dagger_F(b)\right)\left[ s(b,\bar b)-s(b,\bar b+\xi_1\bar\gamma)\right]\nonumber\\
&&
-
A^\lambda(-\bar\gamma)A^{\lambda}(-\gamma) \frac{N_c^2}{2}{\bar r}^i{\bar r}^j\frac{1}{N_c}\tr\left(\partial^iS_F({\bar b})\partial^jS^\dagger_F({\bar b})\right)\left[s(b,{\bar b})-s(b+\xi_1\gamma,\bar b)\right].
\eeq
\\
\noindent (ii) {\it $A^\lambda(-\bar\gamma) A^\lambda(-\gamma)$ term}: The expansion of the $A^\lambda(-\bar\gamma) A^\lambda(-\gamma)$ term can be performed in a similar manner and the result reads
\beq
\label{AA_full}
&&
\hspace{0cm}
A^\lambda(-\bar\gamma)A^{\lambda}(-\gamma)\frac{N_c^2-1}{2}\frac{1}{N_c}\tr\Bigg\{ \Big(S_F(b)-S_F({b+\xi_1\gamma})\Big)\Big(S^\dagger_F({\bar b})-S^\dagger_F({\bar b+\xi_1\bar \gamma})\Big)\nonumber\\
&&
+\frac{1}{2}(1-2\xi_2)\bigg[r^i\Big(\partial^iS_F(b)-\partial^iS_F({b+\xi_1\gamma})\Big)
\Big(S^\dagger_F({\bar b})-S^\dagger_F({\bar b+\xi_1\bar \gamma})\Big)\nonumber\\
&&
\hspace{3cm}
+\bar r^i \Big(S_F(b)-S_F({b+\xi_1\gamma})\Big)
\Big(\partial^iS^\dagger_F({\bar b})-\partial^iS^\dagger_F({\bar b+\xi_1\bar \gamma})\Big)\bigg]\nonumber\\
&&
+\frac{1}{8}(1-2\xi_2)^2\bigg[r^ir^j\Big(\partial^i\partial^jS_F(b)-\partial^i\partial^jS_F({b+\xi_1\gamma})\Big)
\Big(S^\dagger_F({\bar b})-S^\dagger_F({\bar b+\xi_1\bar \gamma})\Big)\\
&&
\hspace{1.5cm}
+
\bar r^i \bar r^j \Big(S_F(b)-S_F({b+\xi_1\gamma})\Big)
\Big(\partial^i \partial^j S^\dagger_F({\bar b})-\partial^i \partial^j S^\dagger_F({\bar b+\xi_1\bar \gamma})\Big)\nonumber\\
&&
\hspace{1.5cm}
+2r^i\bar r^j\Big(\partial^iS_F(b)-\partial^iS_F({b+\xi_1\gamma})\Big)
\Big(\partial^jS^\dagger_F({\bar b})-\partial^jS^\dagger_F({\bar b+\xi_1\bar \gamma})\Big)\bigg]\Bigg\}.
\nonumber
\eeq 
\\
\noindent (iii) {\it ${\cal A}_{\xi_2}^\lambda(-\bar\gamma) A^\lambda(-\gamma)$ term}: The result of the expansion of ${\cal A}_{\xi_2}^\lambda(-\bar\gamma) A^\lambda(-\gamma)$ reads
\beq
\label{tildeAA_full}
&&
\hspace{0cm}
A^\lambda(-\bar\gamma)A^{\lambda}(-\gamma)
\frac{1}{N_c}\tr\Bigg\{ 
\frac{N_c^2-1}{2}\bigg[1-\frac{\xi_2\bar\xi_2}{\xi_1}\frac{\bar r^2}{\bar\gamma^2}\bigg]
\Big(S_F(b)-S_F({b+\xi_1\gamma})\Big)\Big(S^\dagger_F({\bar b})-S^\dagger_F({\bar b+\xi_1\bar \gamma})\Big)\nonumber\\
&&
\hspace{0.5cm}
+\frac{N_c^2-1}{2}\frac{1}{2}r^i(1-2\xi_2)\Big(\partial^iS_F(b)-\partial^iS_F({b+\xi_1\gamma})\Big)\Big(S^\dagger_F({\bar b})-S^\dagger_F({\bar b+\xi_1\bar \gamma})\Big)\nonumber\\
&&
\hspace{0.5cm}
+\frac{1}{2}\bar r^i\bigg[\frac{N_c^2+1}{2}\Big(S_F(b)-S_F({b+\xi_1\gamma})\Big)\partial^iS^\dagger_F({\bar b})\nonumber\\
&&
\hspace{1.5cm}
-\frac{N_c^2-1}{2}(1-2\xi_2)\Big(S_F(b)-S_F({b+\xi_1\gamma})\Big)\partial^iS^\dagger_F({\bar b+\xi_1\bar\gamma})\bigg]\nonumber\\
&&
\hspace{0.5cm}
+\frac{N_c^2-1}{2}\frac{1}{8}r^ir^j(1-2\xi_2)^2\Big(\partial^i\partial^jS_F(b)-\partial^i\partial^jS_F({b+\xi_1\gamma})\Big)
\Big(S^\dagger_F({\bar b})-S^\dagger_F({\bar b+\xi_1\bar \gamma})\Big)\nonumber\\
&&
\hspace{0.5cm}
+\frac{N_c^2-1}{2}\frac{1}{8}\bar r^i\bar r^j\Big(S_F(b)-S_F({b+\xi_1\gamma})\Big)
\Big[\partial^i\partial^jS^\dagger_F({\bar b})-(1-2\xi_2)^2\partial^i\partial^jS^\dagger_F({\bar b+\xi_1\bar \gamma})\Big]\nonumber\\
&&
\hspace{0.5cm}
+\frac{N_c^2+1}{2}\frac{1}{4}r^i\bar r^j (1-2\xi_2)\Big(\partial^iS_F(b)-\partial^iS_F({b+\xi_1\gamma})\Big)\partial^jS^\dagger_F({\bar b})
\nonumber\\
&&
\hspace{0.5cm}
-
\frac{N_c^2-1}{2} \frac{1}{4}r^i\bar r^j (1-2\xi_2)^2\Big(\partial^iS_F(b)-\partial^iS_F({b+\xi_1\gamma})\Big)\partial^jS^\dagger_F({\bar b+\xi_1\bar\gamma})\Bigg\}\nonumber\\
&&
-
A^\lambda(-\bar\gamma)A^{\lambda}(-\gamma) \frac{N_c^2}{2}{\bar r}^i{\bar r}^j\frac{1}{N_c}\tr\left(\partial^iS_F({\bar b})\partial^jS^\dagger_F({\bar b})\right)\left[s(b,{\bar b})-s(b+\xi_1\gamma,\bar b)\right].
\eeq
\\
\noindent (iv) {\it $ A^\lambda(-\bar\gamma){\cal A}_{\xi_2}^\lambda(-\gamma)$ term}: The result of the expansion of $ A^\lambda(-\bar\gamma){\cal A}_{\xi_2}^\lambda(-\gamma)$ reads
\beq
\label{AtildeA_full}
&&
\hspace{0cm}
A^\lambda(-\bar\gamma)A^{\lambda}(-\gamma)
\frac{1}{N_c}\tr\Bigg\{ 
\frac{N_c^2-1}{2}\bigg[1-\frac{\xi_2\bar\xi_2}{\xi_1}\frac{ r^2}{\gamma^2}\bigg]
\Big(S_F(b)-S_F({b+\xi_1\gamma})\Big)\Big(S^\dagger_F({\bar b})-S^\dagger_F({\bar b+\xi_1\bar \gamma})\Big)\nonumber\\
&&
\hspace{0.5cm}
+\frac{1}{2}r^i\bigg[\frac{N_c^2+1}{2}\partial^iS_F(b)-\frac{N_c^2-1}{2}(1-2\xi_2)\partial^iS_F({b+\xi_1\gamma})\bigg]
\Big(S^\dagger_F({\bar b})-S^\dagger_F({\bar b+\xi_1\bar\gamma})\Big)\nonumber\\
&&
\hspace{0.5cm}
+\frac{N_c^2-1}{2}\frac{1}{2}{\bar r}^i(1-2\xi_2)\Big(S_F(b)-S_F({b+\xi_1\gamma})\Big)\Big(\partial^iS^\dagger_F({\bar b})-\partial^iS^\dagger_F({\bar b+\xi_1\bar \gamma})\Big)\nonumber\\
&&
\hspace{0.5cm}
+\frac{N_c^2-1}{2}\frac{1}{8}r^ir^j\Big(\partial^i\partial^jS_F(b)-(1-2\xi_2)^2\partial^i\partial^jS_F({b+\xi_1\gamma})\Big)
\Big(S^\dagger_F({\bar b})-S^\dagger_F({\bar b+\xi_1\bar \gamma})\Big)\nonumber\\
&&
\hspace{0.5cm}
+\frac{N_c^2-1}{2}\frac{1}{8}\bar r^i\bar r^j (1-2\xi_2)^2\Big(S_F(b)-S_F({b+\xi_1\gamma})\Big)
\Big(\partial^i\partial^jS^\dagger_F({\bar b})-\partial^i\partial^jS^\dagger_F({\bar b+\xi_1\bar \gamma})\Big)\nonumber\\
&&
\hspace{0.5cm}
+\frac{N_c^2+1}{2}\frac{1}{4}r^i\bar r^j (1-2\xi_2) \partial^iS_F(b)  \Big(\partial^jS^\dagger_F({\bar b})-\partial^jS^\dagger_F({\bar b+\xi_1\bar \gamma})\Big)
\nonumber\\
&&
\hspace{0.5cm}
-
\frac{N_c^2-1}{2} \frac{1}{4}r^i\bar r^j (1-2\xi_2)^2 \partial^iS_F({b+\xi_1\gamma}) \Big(\partial^jS^\dagger_F({\bar b})- \partial^jS^\dagger_F({\bar b+\xi_1\bar\gamma}) \Big) \Bigg\}\nonumber\\
&&
-A^\lambda(-\bar\gamma)A^{\lambda}(-\gamma) \frac{N_c^2}{2}r^ir^j\frac{1}{N_c}\tr\left(\partial^iS_F(b)\partial^jS^\dagger_F(b)\right)\left[ s(b,\bar b)-s(b,\bar b+\xi_1\bar\gamma)\right].
\eeq

Now, we can combine all the terms using Eqs.~\eqref{tildeAtildeA_full}-\eqref{AtildeA_full}. Then we see that the first non-vanishing terms are ${\cal O}(r\bar r)$ and the result reads
\beq
\label{Full_after_expansion}
&&
\hspace{-1.5cm}
A^\lambda(-\bar\gamma)A^{\lambda}(-\gamma)\frac{N_c^2-1}{2}r^i\bar r^j
\Bigg\{
\bigg[ \xi_2^2-\frac{(1-2\xi_2)}{N_c^2-1}\bigg]\frac{1}{N_c}\tr\Big(\partial^iS_F(b)\partial^jS^\dagger_F({\bar b})\Big)\nonumber\\
&&
\hspace{3cm}
-
\frac{N_c^2}{N_c^2-1}\frac{1}{N_c}\tr\Big(\partial^iS_F(b)S^\dagger_F({\bar b})\partial^jS_F({\bar b})S^\dagger_F(b)\Big)s(b,\bar b)\Bigg\}\, .
\eeq


\end{document}